\documentclass{iopart}
\expandafter\let\csname equation*\endcsname\relax 
\expandafter\let\csname endequation*\endcsname\relax 
\usepackage{amsmath}
\usepackage{amssymb}
\usepackage{graphicx}% Include  files
\usepackage{bm}% bold math
\usepackage{epstopdf}
\usepackage{color}
\usepackage[colorlinks=true, pdfstartview=FitV, linkcolor=blue, citecolor=blue, urlcolor=blue]{hyperref}
\definecolor{shadecolor}{rgb}{0.95, 0.95, 0.86}

\newcommand{\bt}{\beta}

\newcommand{\part}{\partial}

\def\bt{\begin{theorem}}
\def\et{\end{theorem}}
\def\bc{\begin{corollary}}
\def\ec{\end{corollary}}
\def\bx{\begin{example}\small}
\def\ex{\end{example}}
\def\bxr{\begin{exercise}\small}
\def\exr{\end{exercise}}
\def\bl{\begin{lemma}}
\def\el{\end{lemma}}
\def\bd{\begin{definition}}
\def\ed{\end{definition}}
\def\bp{\begin{proposition}}
\def\ep{\end{proposition}}

\def\br{\begin{remark}}
\def\er{\end{remark}}

\def\be{\begin{equation}}
\def\ee{\end{equation}}
\def\&{\hspace{-15pt}&}
\def\bea{\begin{eqnarray}}
\def\eea{\end{eqnarray}}

\def \part{\partial}

\def\1{{\bf 1}}

\begin{document}

\title[Dispersive shock waves in BO type systems ]{Dispersive shock waves in systems with nonlocal 
dispersion of Benjamin-Ono type}

\author{G. A. El$^{1}$,  L.T.K. Nguyen $^{2}$, and  N. F. Smyth $^{3,4}$}

\address{
$^{1}$ Centre for Nonlinear Mathematics and Applications, \\
Department of Mathematical Sciences, Loughborough University, \\ Loughborough LE11 3TU, U.K.} 
\ead{g.el@lboro.ac.uk}

\address{$^{2}$ Institute of Applied Mechanics, University of Stuttgart,\\
Pfaffenwaldring 7, 70569 Stuttgart, Germany}
 \ead{nguyen@mechbau.uni-stuttgart.de}      
 
 \address{$^{3}$ School of Mathematics, University of Edinburgh,\\
Edinburgh, Scotland,  EH9 3FD, U.K. \\

{$^4$} School of Mathematics and Applied Statistics, University of Wollongong, \\
Northfields Avenue, Wollongong, New South Wales, 2522, Australia}
\ead{N.Smyth@ed.ac.uk}
\date{\today}

\begin{abstract}
We develop a general approach to the description of dispersive shock waves (DSWs)
for a class of nonlinear wave equations with a nonlocal Benjamin-Ono type dispersion term involving the 
Hilbert transform.  Integrability of the governing equation is not a pre-requisite for the application of 
this method which represents a modification of the DSW fitting method previously developed for  
dispersive-hydrodynamic systems of Korteweg-de Vries (KdV) type (i.e.\ reducible to the KdV equation in the  
weakly nonlinear, long wave, unidirectional approximation).  The developed method is applied to the 
Calogero-Sutherland dispersive hydrodynamics for which the classification of all solution types arising from
the Riemann step problem is constructed and the key physical parameters (DSW edge speeds, lead soliton amplitude, 
intermediate shelf level) of all but one solution type are obtained in terms of the initial step data. 
The analytical results are shown to be in excellent agreement with results of direct numerical simulations.
\end{abstract}

%\submitto{\NL}

\section{Introduction}
\label{sec:intro}

The most well known solution of nonlinear dispersive wave equations is the solitary wave,
termed a soliton for integrable equations \cite{ablowitz}.  Another
generic solution of such equations is the dispersive shock wave (DSW), also known as an undular
bore in fluids applications \cite{borereview}.  The major difference between these two fundamental types of 
solution is that a solitary wave is a steady, propagating solution, whilst a typical DSW is an inherently unsteady   
wavetrain exhibiting a solitary wave at one edge, while at the other edge it degenerates into a linear wavepacket 
propagating with the group velocity.  The relative positions of the solitary wave and linear edges determine the DSW 
orientation \cite{borereview}.  Another general characteristic of a DSW is its polarity, defined by the polarity 
of the solitary wave at one of its edges (waves of elevation are associated with positive polarity). 
%The polarity and the orientation of a DSW generally are two independent parameters determined by the signs of the coefficients for nonlinear and dispersion terms in the governing equation although  for some systems, such as the Korteweg -- de Vries (KdV)  equation they both can be defined by a sign of a  product of these coefficients \cite{borereview1, borereview2}
In between the solitary wave and linear edges, a ``classical'' DSW consists of a modulated periodic wave, with the
modulation in amplitude, wavelength and mean height giving a smooth transition between the solitary wave and 
linear wave edges.  
%In convex dispersive-hydrodynamic system the DSW polarity and orientation are rigidly connected with each other \cite{borereview}.

DSWs have been observed in a wide range of physical applications.  They classically
arise as tidal bores in coastal regions with strong tides and suitable coastal topography which acts
to funnel the tide, examples being the Severn River in England, the Bay of Fundy in Canada and the 
Araguari River in Brazil.  Undular bores have also been observed as internal waves in the atmosphere 
\cite{clarke} and in the semi-diurnal internal tide \cite{nwshelf}.  DSWs have also been proposed as models for 
unsteady processes in viscously deformable media, e.g.\ in magma migration through the mantle or in the 
buoyant ascent of a low density fluid through a viscously deformable conduit \cite{hoefer_conduit,scott1}.
Recently, DSWs have been experimentally observed in nonlinear optical media, such as photorefractive crystals 
\cite{fleischer2,fleischer}, nonlinear thermal optical media \cite{trillo6,boreexp} and nonlinear
optical fibres \cite{trilloresfour,trilloresnature,trillo7, trillo2017}.  

From the mathematical modelling viewpoint the best known example of a DSW is the solution of the KdV equation 
with step initial data (the so-called dispersive Riemann problem). This solution was first obtained by Gurevich 
and Pitaevskii \cite{gur} (see also \cite{bengt}) using the Whitham modulation equations for the KdV 
equation \cite{modproc} (see also \cite{kamch_book}).  The subsequent development of rigorous DSW theory based 
on the inverse scattering transform (see \cite{miller_review} and references therein)
%\cite{kotlyarov}, \cite{lax_lev}, \cite{ven}, \cite{dvz},\cite{teschl}, \cite{grava}, 
have confirmed the Gurevich-Pitaevskii modulation solution as the essential part of the full, long time asymptotic 
solution of the KdV step problem.  At the same time, one should stress that, while being consistent with the 
rigorous theory for integrable systems, the modulation theory approach to DSW description is not based on 
integrability and, hence, has the advantage of applicability to a broader class of physically relevant dispersive 
equations under certain, relatively mild structural assumptions \cite{borereview,el1, hoefer_euler}.

The KdV DSW structure is typical for convex dispersive-hydrodynamic systems \cite{borereview, siam_review},
exhibiting a linear dispersion relation with a generic long wave expansion of the form (uni-directional 
propagation is assumed for simplicity)
\begin{equation}
\omega = k V({\bar u})  - \mu k^3+ o(k^3), \ \  0< k \ll 1, \quad \mu \ne 0,
\end{equation}
where $V({\bar u}) \ne 0$ is the long-wave characteristic speed ($V(\bar u) = \bar u $ for the KdV equation)
and $\bar{u}$ is the mean level of $u$.   If $V({\bar u})= 0$ at some ${\bar u}=\bar{u}_c$, as for the mKdV or 
Gardner equations exhibiting a non-convex hyperbolic flux, non-KdV type DSWs (also termed ``non-classical DSWs''
\cite{borereview, siam_review}) become possible, which include contact (trigonometric) DSWs and kinks 
(non-oscillating, undercompressive DSWs), see \cite{siam_review,gardner,marchant_mkdv}.  Other types of 
non-classical DSWs are the radiating DSWs found for the Kawahara equation \cite{hoefer_kawahara} and for the 
NLS equation with higher order dispersion \cite{trillores} or non-local nonlinearity \cite{nemboreel,nembore}. 

In this work we shall consider DSWs in a system with non-local dispersion of Benjamin-Ono (BO) type exhibiting,
in the uni-directional case, a linear dispersion relation of the form
\begin{equation}\label{bo_disp}
\omega = k V({\bar u})  + \sigma k |k| ,   \quad  \sigma \ne 0.
\end{equation}
The BO equation itself has the form \cite{benjamin,ono}
\begin{equation}
 \frac{\partial u}{\partial t} + 2u\frac{\partial u}{\partial x}
 +   \mathcal{H}[u_{xx}] = 0,
 \label{e:bo}
\end{equation}
where 
\begin{equation}
\mathcal{H}[f] =  \mbox{P.V.} \frac{1}{\pi} \int_{-\infty}^{\infty} \frac{f(y)}
 {y-x} \: dy 
 \end{equation}
is the Hilbert transform of $f(x)$ and $\mbox{P.V.}$ denotes the principal value.  The linear dispersion relation 
for the BO equation (\ref{e:bo}) has the form (\ref{bo_disp}) with $V({\bar u})= 2 \bar u$ and $\sigma =-1$.  
The BO equation governs weakly nonlinear waves in a layer of stratified fluid overlaying  a deep layer of weakly 
stratified fluid \cite{benjamin,ono}, in contrast to the Korteweg-de Vries (KdV) equation which governs weakly 
nonlinear shallow water waves.  In particular, the BO equation models the atmospheric phenomenon called a morning 
glory, which is an atmospheric undular bore \cite{clarke,anne}.  A very recent application of the BO equation is the 
spectral dynamics of incoherent shocks in nonlinear optics \cite{BO_trillo}.

The key to deriving the DSW solution of the KdV equation \cite{gur} was the ability
to set the KdV modulation equations in Riemann invariant form \cite{modproc}.  It was subsequently shown 
that this property of the KdV modulation equations is a consequence of the integrability of the KdV equation 
\cite{flash,lax_lev}.
%
% integrability of
%the KdV equation means that (i) its (multi-)phase modulation equations can be derived via a standard procedure
%applicable to other integrable equations and (ii) integrability guarantees that these hyperbolic modulation equations 
%can be set in Riemann invariant form \cite{flash}.  
This connection between integrability and the structure of the modulation equations then enabled DSW solutions to 
be found for other integrable equations, such as the nonlinear Schr\"odinger (NLS) equation \cite{el_krylov95,elnls,gur_kryl}, 
the Kaup-Boussinesq equation \cite{kb_sapm2,kb_sapm1}, the modified KdV (mKdV) equation \cite{siam_review,marchant_mkdv}, 
the Gardner equation \cite{gardner} and the Sine-Gordon equation \cite{gershenzon, sgbore}.  The BO equation is also 
integrable and its modulation equations in Riemann invariant form were first derived by \cite{dob} and then used for 
the description of the BO DSW for the step problem \cite{tim,matsuno1} and also for arbitrary hump-like initial 
data \cite{matsuno2}.   Rigorous analysis of the semi-classical limit of the BO equation has confirmed the modulation 
theory results \cite{miller1, miller2}.  Also, the modulation DSW theory for the BO equation modified by a small 
viscous term was constructed in \cite{matsuno3}.

The above developments of DSW modulation theory are based on the availability of the Riemann invariant form for 
the modulation equations.  This then raises the question of the determination of DSW solutions of non-integrable,
nonlinear wave equations for which the Riemann invariant form of the modulation equations is generally not available.  
%The derivation of modulation equations for non-integrable nonlinear
%wave equations is, in general, not straightforward and even if these can be determined and they are hyperbolic,
%setting these in Riemann invariant form is not guaranteed as this is a form of Pfaff's problem.  The determination of DSW solutions of non-integrable nonlinear wave equations is then an open problem.  
A partial solution to this was found in \cite{el2,el1} for which a method was developed to determine the closure 
conditions for the dispersive regularisation of an initial step without solving the full modulation equations, but 
instead by directly deriving the DSW locus and the speeds of its edges.  Due to the inherently unsteady structure 
of DSWs these conditions fundamentally differ from the traditional Rankine-Hugoniot closure describing the steady, 
travelling wave transitions typical for viscous and diffusive-dispersive shocks \cite{siam_review,whitham}.  Unlike 
classical shocks, DSWs are characterised by two speeds of propagation describing the motion of their solitary wave 
and harmonic (linear) wave edges.  The DSW fitting method \cite{el2} is based on the observation that at the edges 
of a DSW the Whitham modulation equations always have a degenerate form for any nonlinear wave equation whose periodic 
travelling wave solution is determined by a differential equation of the form 
\begin{equation}
 u_{\theta}^{2} = Q(u)r^2(u),
\label{e:kdvode}
\end{equation}
where $\theta = kx - \omega t$ is the phase, with $k$ the wavenumber and $\omega$ the frequency, and
$Q$ is a cubic polynomial in $u$ with roots at $u=u_j$, $j=1,2,3$.  The function $r(u)$ is some smooth function 
which does not vanish at $u=u_j$ \cite{el2,borereview,el1,hoefer_conduit}.  The DSW structure for such a nonlinear 
wave equation is the same as that for the KdV DSW \cite{el2,borereview,el1}.  The DSW solutions of the  
NLS \cite{elnls}, 
%Sine-Gordon \cite{sineg},  
conduit \cite{hoefer_conduit}, photo-refractive \cite{hnls,elphoto} and optical colloid equations 
\cite{colloid} are of this general form, the last three being non-integrable equations.  

The universal structure of the degenerate Whitham modulation equations in the zero amplitude (linear wave) and 
zero wavenumber (solitary wave) limits was shown in \cite{el2,el1} to enable the determination of the leading and 
trailing edges of DSWs of KdV type.  However, as already mentioned, not all DSWs are of KdV type due to 
the fundamentally different structure of the dispersive term, so that the steady travelling wave solution is 
not necessarily determined by an equation of the form (\ref{e:kdvode}).  The BO equation is a prominent example 
of such a non-KdV type model equation whose soliton solutions exhibit algebraic decay, in contrast to KdV 
solitons which decay exponentially.  This implies that the DSW fitting method in its original form 
\cite{el2,el1} (see also \cite{borereview}) cannot be applied to determine the soliton edge of a BO DSW.  

In this work the DSW fitting method \cite{el2, el1} will be extended to cover the derivation of the 
leading and trailing edges of DSWs and the classification of the Riemann problem solutions for nonlinear wave 
equations with BO type dispersion.  As the BO equation is integrable \cite{dob} and so has a known full DSW modulation 
solution \cite{tim}, the solution method for the leading and trailing edges will be verified with this known 
DSW solution.  Once the solution method is verified, it will be applied to characterise the DSW solutions of the  
Calogero-Sutherland (CS) dispersive hydrodynamics arising in the continuous, thermodynamic limit of the original 
CS model governing quantum many-body systems \cite{cal2,suth1,suth2}.  The equations for a CS ``fluid'' with  
density $\rho$ and velocity $v$ have the form \cite{abanov2,abanov1,polychronakos}
\begin{eqnarray}
&& \frac{\partial \rho}{\partial t} + \frac{\partial}{\partial x} (\rho v)  =  0, \label{e:mass1} \\
&& \frac{\partial v}{\partial t} + \frac{\partial}{\partial x} \left[ \frac{1}{2}v^{2} + \frac{1}{2}
\pi^{2} g^{2} \rho^{2} -  g^2 \left( \frac{1}{4} \frac{\rho_{xx}}{\rho} - \frac{1}{8} \frac{\rho_{x}^{2}}
{\rho^{2}}  - \pi  \mathcal{H}[\rho_{x} ] \right)\right]  =  0,  \label{e:mom1}
\end{eqnarray}
where $g>0$ is the coupling constant \cite{abanov1,polychronakos}.  Equations (\ref{e:mass1}) and (\ref{e:mom1})  
have the form of Eulerian dispersive hydrodynamics \cite{borereview, siam_review,hoefer} with the equation of 
state $P(\rho) \sim \rho^3$ complemented by dispersive terms of two different types.  One is of the nonlinear 
Schr\"odinger type, while the other one is of BO type.  Introducing a complex function 
$\psi=\rho^{1/2}\exp{(\frac{i}{g} \int v dx)}$, we obtain an equivalent form of system (\ref{e:mass1}) and 
(\ref{e:mom1}) \cite{abanov2}
\begin{equation}\label{2BO_compl}
\frac{i}{g} \psi_t = \left[ -\frac12 \partial_x^2 + \frac{\pi^2}{2}|\psi|^4 + 
\pi \mathcal{H}[\partial_x |\psi|^2 ]\right] \psi \, .
\end{equation}
An equivalence of the CS dispersive hydrodynamics (\ref{e:mass1}) and (\ref{e:mom1}) to the so-called intermediate nonlinear
Schr\"odinger equation \cite{pelin} describing the evolution of a modulated internal wave in a deep stratified fluid 
has been shown in Ref.\ \cite{abanov2}.

The linear dispersion relation for the CS system (\ref{e:mass1}) and (\ref{e:mom1}) (or, equivalently, 
eq.\ (\ref{2BO_compl})) obtained by linearisation about a background state $\rho= \bar \rho$ and $v= \bar v$ has 
the form 
\begin{equation} \label{dr_2BO}
\omega = kV_{\pm} (\bar{\rho}, \bar{v}) \mp \frac{1}{2}gk|k|,
\end{equation}
where $V_{\pm} = \bar{v} \pm \pi g \bar{\rho}$ are the characteristic velocities of the CS system (\ref{e:mass1}) 
and (\ref{e:mom1}) in the dispersionless, long wave limit.  One can see that both branches of the dispersion 
relation (\ref{dr_2BO}) have the form (\ref{bo_disp}) characteristic of BO type equations.  Indeed, 
equation (\ref{2BO_compl}) has been termed a bi-directional BO equation \cite{abanov2}.

The CS dispersive hydrodynamic equations (\ref{e:mass1}) and (\ref{e:mom1}) (in what follows ``the CS equations'') 
also form an integrable system, but their Whitham modulation equations have not been derived to our knowledge.  
Although such a derivation is clearly possible in the framework of finite-gap averaging theory following the route 
of \cite{dob,flash}  we, instead, shall take advantage of an appropriate modification of the DSW fitting 
method \cite{el2,  borereview} to directly evaluate the key parameters of DSWs generated in the CS Riemann 
problem using just the linear dispersive relation (\ref{dr_2BO}), thus bypassing the determination of the full 
modulation solution.  Our analysis also offers a straightforward generalisation to other types of nonlinearities 
by assuming an arbitrary convex pressure law $P =f(\rho) $ for which the (generalised) CS system is likely to be 
non-integrable.  This natural generalisation can be viewed as a counterpart of the extension of the integrable 
NLS DSW theory \cite{elnls,gur_kryl} to the case of the generalised NLS equation \cite{elphoto,hoefer_euler} 
having applications in nonlinear optics and cold atom physics.

In this paper we produce a classification of solutions of the Riemann problem for CS dispersive hydrodynamics, 
which contains six possible solution forms, each representing a certain combination of two waves, DSWs and/or 
rarefaction waves, separated by a constant shelf.   Five of the configurations are described analytically using 
the modified DSW fitting method \cite{el2,  borereview}.  The sixth involves the interaction of two DSWs, producing an 
oscillating intermediate shelf similar to the analogous case for the defocusing NLS Riemann problem \cite{elnls}. 
This sixth configuration requires a knowledge of the full modulation equations, and hence cannot be described by the 
DSW fitting method.  The five available modulation solutions are compared with full numerical solutions of the CS 
dispersive hydrodynamic equations and excellent agreement is found. 

\section{Benjamin-Ono DSW}
\label{s:bo}

In this section the leading, solitary wave, and the trailing, linear wave, edges of the DSW solution of the BO
equation (\ref{e:bo}) will be derived using the DSW fitting method \cite{el2, borereview} extended to nonlinear wave 
equations with BO type dispersion.  The modulation solution describing the BO DSW is well known \cite{tim,matsuno1}. 
In the present paper it will be used to verify the developed method which will then be applied to the CS system 
(\ref{e:mass1}) and (\ref{e:mom1}) for which the DSW solution is not known.

To generate a DSW the step initial condition 
\begin{equation}
u = \left\{ \begin{array}{cc}
               u_{-}, & x < 0, \\
               u_{+}, & x > 0 
            \end{array}
    \right.
\label{e:boic}    
\end{equation}               
will be assumed.  Here, $u_{-}$ and $u_{+}$ are constants and $u_{+} < u_{-}$ for DSW formation. 

\subsection{Integrable theory}
\label{sec:int_theory}

The BO equation is one of a class of integrable nonlinear wave equations which can be solved 
via the method of inverse scattering \cite{ablowitz,novikov}.  As a result, the Whitham modulation equations  
for the slowly varying periodic wave solution of the BO equation can be determined in Riemann invariant form 
\cite{dob}.  
%For the KdV equation the Riemann invariants of the modulation system were found by Whitham in the single-phase case \cite{modproc}. The general method of finding Riemann invariants using the integrability via the inverse scattering was developed  in \cite{flash}.  This method was applied to the case of the BO equation in \cite{dob}. 
Remarkably, unlike in the case of the KdV modulation system \cite{flash,modproc}, whose characteristic velocities 
contain rather involved combinations of (hyper)elliptic integrals depending on all the Riemann invariants, the BO 
$N$-phase modulation system is a system of $2N+1$ {\it decoupled} simple wave equations
\begin{equation}
\frac{\partial r_i}{\partial t} + 2 r_i\frac{\partial r_i}{\partial x} = 0, \quad i=1, \dots, 2N+1 \, ,
\label{e:modulation1}
\end{equation}
where the Riemann invariants $r_1 \le r_2 \le \dots \le r_{2N+1}$ are uniquely related to the $2N+1$ parameters  
characterising the finite gap, multiphase BO solution \cite{dob}.   Each of the equations (\ref{e:modulation1}) has 
the same form as the dispersionless limit of the BO equation (\ref{e:bo}).  The BO Riemann step problem involves 
only the single phase solution, $N=1$.  We note that for $N=1$ the BO travelling wave solution is
expressed in terms of trigonometric functions \cite{benjamin} and the BO modulation equations in Riemann invariant form 
can be readily found from the Whitham system obtained from the averaged variational principle \cite{matsuno1, matsuno2}.
%\begin{equation}
%\frac{\partial u}{\partial t} + 2 u \frac{\partial u}{\partial x} = 0.
%\label{e:displess_bo}
%\end{equation}

When the Riemann invariants of the single phase modulation system are known, the corresponding DSW solution is 
readily determined as a simple wave solution of these modulation equations \cite{borereview,bengt,gur}.  The BO 
DSW modulation solution in the form of a centred rarefaction fan has been found as a self-similar solution of the 
system (\ref{e:modulation1}) for $N=1$ in \cite{tim,matsuno1}.  It has the form 
\begin{equation}
 r_1= u_+, \quad r_3=u_-, \quad r_2=x/(2t)\, .
 \end{equation}
This modulation solution, after insertion into the BO travelling wave solution, describes a DSW with a positive 
polarity and orientation \cite{borereview}, exhibiting a leading soliton edge and a trailing harmonic edge.  The BO 
DSW is characterised by two speeds, $s_+ = 2u_- $ and $s_-=2u_+$, the speeds of the leading (soliton) and the 
trailing (linear) edges, respectively.  The BO soliton riding on the background $u= {\bar u}$
is characterised by the speed-amplitude relation $c_s= 2\bar{u} + a/2$.  So using $\bar u = u_+$ and $c_s=s_+$ 
we find the amplitude of the algebraic soliton at the leading edge of the DSW, $a_+= 4 \Delta$, where 
$\Delta=u_- - u_+$ is the initial jump.  

We note that the outlined modulation solution does not yield the exact phase of the DSW wavetrain.  The determination 
of the DSW phase involves a more delicate analysis based on the Riemann-Hilbert problem approach to semi-classical 
inverse scattering \cite{miller_review}.
%This known DSW solution will be used to verify the extension of the method of the DSW fitting \cite{el1,el2} from equations with KdV type dispersion to equations with BO type dispersion.

%Outside of the DSW dispersion plays a negligible role and the solution is governed by the dispersionless limit (\ref{e:nondispbo}) of the BO equation (\ref{e:bo}).  
%This equation  can be set in characteristic form as
%\begin{equation}
%\frac{du}{dt} = 0 \quad \mbox{on} \quad \frac{dx}{dt} = V = 2u .
%\label{e:charnondispbo}
%\end{equation}

\subsection{DSW fitting}
\label{sec_DSW_fitting}

Here we develop a BO modification of the DSW fitting method \cite{el2,borereview,el1} which has so far been applied to
equations of KdV type.  This method will enable us to determine the leading and trailing edge speeds of a BO 
type DSW using just the linear dispersion relation.  Since integrability is not a pre-requisite for the DSW fitting 
method we shall consider the generalised BO (gBO) equation 
\begin{equation}
 \frac{\partial u}{\partial t} + V(u)\frac{\partial u}{\partial x}
 -   \sigma \mathcal{H}[u_{xx}] = 0,
 \label{e:gbo}
\end{equation}
where $V'( u) \ne 0 $ and $\sigma \ne 0$.  The well-posedness of solutions of the gBO equation (\ref{e:gbo}) with 
$V(u) \sim u^\gamma$ and $\sigma = -1$ was studied in \cite{bona}. 

The gBO equation (\ref{e:gbo}) is characterised by the linear dispersion relation (\ref{bo_disp}), so that the group 
velocity is
%Seeking harmonic
%wave solutions  of the linearised BO equation in  the form
%\begin{equation}
%u = \bar{u} + Ae^{i(kx - \omega t)}
%\label{e:wavebo}
%\end{equation}
%where $|A| \ll \bar{u}$, we obtain the linear dispersion relation
%\begin{equation}
%\omega = 2\bar{u}k - k^{2} .
%\label{e:dispbo}
%\end{equation}
%assuming $k \ge 0$.
\begin{equation}
 c_{g} = \frac{\partial \omega}{\partial k} = V(\bar u) + 2 \sigma k,
\label{e:grouplin}
\end{equation}
where we have omitted the modulus sign for $|k|$ on assuming $k \ge 0$.  The polarity $p$ and orientation $d$ of 
a DSW for the gBO equation (\ref{e:gbo}) are determined as \cite{borereview,siam_review}
\begin{equation}\label{e:pol_or}
p=  - \hbox{sgn}{\{ V'( u) \sigma\} }, \quad d =  - \hbox{sgn} \  \sigma.
\end{equation}
If $V'(u)>0$, we have $p=d$ (we recall that $d=1$ means that the soliton edge is the DSW leading edge). 
For the BO equation (\ref{e:bo}) we have $\sigma =-1$ and $V'(u)=2$, so that $p_{BO}=d_{BO}=1$.

The essence of the DSW fitting method \cite{el2,borereview,el1} to determine the edges of a DSW in systems with KdV type 
structure is the realisation that these trailing and leading edges are characteristics of the modulation equations 
evaluated in two distinguished limits: the zero-amplitude, harmonic limit $a \to 0$ and the zero wavenumber, 
solitary wave limit $k \to 0$.  This imposes certain restrictions (characteristic relations) on the admissible 
values of the modulation parameters at these edges: $k$ and $\bar u$ at the harmonic edge and $a$ and $\bar u$ at 
the solitary wave edge.  Remarkably, these characteristic relations 
in the form of ordinary differential equations can be found directly from the zero amplitude and zero wavenumber 
reductions of the modulation system which are available {\it without knowledge of the full modulation system}.  
One of the important features of the DSW fitting method is the use of a special ``conjugate'' system of modulation 
variables in the zero wavenumber limit which establishes a remarkable symmetry in the characteristic relations 
determining the harmonic and solitary wave edges.  The availability of the conjugate modulation variables relies 
on the special (although rather generic) form of the ordinary differential equation (\ref{e:kdvode}) determining 
the periodic solution, which is typical for KdV-like dispersive hydrodynamics equations.  The BO travelling 
wave equation is not of KdV type, yielding trigonometric, rather than elliptic function, solutions.  As a result,
the harmonic, linear edge of a DSW can be treated in the same way as in the KdV case, but the solitary wave edge exhibits 
algebraic, rather than exponential behaviour, and requires the development of a new approach.

In what follows we shall assume that the gBO equation \eqref{e:gbo} satisfies the pre-requisites for
Whitham modulation theory, i.e.\ it has a family of periodic travelling wave solutions parametrised by three constants 
and possesses at least two conservation laws, see \cite{el2, borereview}. The above properties are, of course, guaranteed 
for the BO equation itself.
 
\subsubsection{Harmonic edge}
\label{sec:BO_harm}

In the zero amplitude limit the modulation system consists of a hyperbolic, dispersionless equation for the mean flow  
complemented by a wave conservation law for linear dispersive waves.  As a consequence of this, it can be shown that 
the characteristic relation between the two modulated wave parameters, the wavenumber $k$ and the mean
height $\bar{u}$, can be reduced to the differential equation 
\cite{el2,borereview,el1}
\begin{equation}
\frac{dk}{d\bar{u}} = \frac{\frac{\partial \omega}{\partial \bar{u}}}
{V(\bar{u}) - \frac{\partial \omega}{\partial k}} ,
\label{e:linbobore}
\end{equation}
where the wave frequency $\omega$ is determined by the linear dispersion relation.  Note that the characteristic
equation (\ref{e:linbobore}) for linear waves  is valid for any type of dispersion as the zero amplitude limit of 
the Whitham modulation equations just depends on the linear dispersion relation and not the details of how this is 
derived \cite{whitham}.  

%The linear edge of the 
%Benjamin-Ono DSW has the same structure as that for the KdV DSW, so this equation will also apply at the linear,
%trailing edge of the Benjamin-Ono DSW.  

Using the dispersion relation (\ref{bo_disp}) for the gBO equation (\ref{e:gbo}), we find that
the characteristic equation (\ref{e:linbobore}) assumes the form
\begin{equation}
\frac{dk}{d\bar{u}} = -\frac{1}{2\sigma} V'(\bar u).
\label{e:linbobore2}
\end{equation}
%This characteristic  relation for the linear edge of the BO type DSW is much simpler than for the KdV type case \cite{el1,el2}.  This is a reflection of the simple nature of the modulation equations (\ref{e:modulation1})  for the BO equation.  
%In contrast, the modulation equations for
%the KdV equation form a third order hyperbolic system with the characteristic velocities given in terms of 
%elliptic integrals \cite{whitham}.  
To apply the characteristic equation (\ref{e:linbobore2}) to the description of the DSW harmonic edge it must be 
supplied with an appropriate initial condition.  To be definite, we assume that $\sigma=-1$ and $V'(u)>0$, as in the 
standard form of the integrable BO equation (\ref{e:bo}).  Then the DSW orientation (\ref{e:pol_or}) gives $d=1$ and  
the DSW solitary wave edge is associated with the right state $u=u_+$.  The initial condition for (\ref{e:linbobore2}) 
is then found from the requirement that $k=0$ at the DSW leading edge where $\bar{u} = u_{+}$ (see 
\cite{el2,borereview} for a detailed explanation of why the solitary wave edge condition for $k$ is 
applicable to the zero amplitude characteristic equation (\ref{e:linbobore})).
%To match with the leading, soliton edge of the DSW, equation (\ref{e:linbobore2})
%for the trailing, linear wave edge of the DSW is solved with the 
%boundary condition $k=0$ at $\bar{u} = u_{+}$ on using the initial
%condition (\ref{e:boic}).  This boundary condition arises as at the leading edge of the 
%bore there is a solitary wave, which has zero wavenumber, on the mean level $u_{+}$.  
The solution of (\ref{e:linbobore2}) is then
\begin{equation}
k(\bar u) = \frac{1}{2} [V(\bar{u}) - V(u_{+})],
\label{e:linwaven}
\end{equation}
which gives the wavenumber at the trailing edge of the DSW as
\begin{equation}
k_-=k(u_-) =  \frac{1}{2} [V(u_-) - V(u_{+})].
\end{equation}
Finally, the velocity $s_-$ of the trailing, linear wave edge of the DSW is determined by the 
group velocity (\ref{e:grouplin}) as
\begin{equation}
s_- = c_{g}(u_-) = V(u_-) - 2k_- = V(u_{+}).
\label{e:grouplinbob}
\end{equation}
In particular, for the case of the integrable BO equation (\ref{e:bo}) with $V(u)=2u$ and $\sigma=-1$ we obtain 
$s_-= 2u_+$, which agrees with the result of \cite{tim,matsuno1} reproduced in Sec.\ \ref{sec:int_theory}.

The case of positive dispersion $\sigma= 1$ is dealt with in the same way taking into account the change in the DSW 
polarity and orientation so that the zero wavenumber initial condition for (\ref{e:linbobore2}) should be 
formulated at $\bar u =u_-$.

\subsubsection{Soliton edge}
\label{sec:sol_edge_BO}

While the harmonic edge of a BO type DSW can be determined using the above method of \cite{el2,borereview,el1},
the same method cannot be used to determine the solitary wave edge.  This is because this method requires the ordinary 
differential equation governing the periodic wave solution to have the form (\ref{e:kdvode}) with the 
``potential curve'' $Q(u)$ exhibiting three real roots, implying the exponential decay of solitary wave solutions.
%\begin{equation}
% u_{\theta}^{2} = P(u),
%\label{e:polyde}
%\end{equation}
%where $\theta = kx - \omega t$ is the phase and $P(\theta)$ is a polynomial.  
%For the KdV equation
%$P = (p-u)(u-q)(u-r)$.  
Due to the Hilbert transform the periodic wave solutions of the gBO equation (\ref{e:gbo}) are not governed by 
an equation of the form (\ref{e:kdvode}), with the consequence that the solitary wave solutions of BO type 
equations do not decay exponentially as $x \to \pm \infty$, but have algebraic decay.  For the integrable BO equation 
(\ref{e:bo}) the solitary wave (soliton) solution is
\begin{equation}
 u = \bar{u} + \frac{a}{1 + \frac{1}{4}a^{2}\left[ x - c_s(\bar{u}, a)t \right]^{2}}, \quad
 c_s= 2\bar{u} + \frac{1}{2}a.
 \label{e:bosol}
\end{equation}
We proceed with the determination of the solitary wave DSW edge for the gBO equation (\ref{e:gbo})
assuming, when necessary, that the speed-amplitude relation $c_s(\bar u, a)$ is known.

The key to the determination of the solitary wave edge by the DSW fitting method for equations of KdV type is 
the use of the conjugate dispersion relation $\tilde \omega (\bar u, \tilde k)$ obtained as 
$\tilde \omega = - i \omega (\bar u, i \tilde k)$, where $\tilde k$ is the conjugate wavenumber, an amplitude 
type variable whose meaning is explained below.  It is not difficult to see that for KdV type systems 
the conjugate dispersion relation is simply the linear dispersion relation for the associated dynamics in the 
$(ix, it)$ plane \cite{el2}.  In terms of the travelling wave equation (\ref{e:kdvode}) this corresponds to changing 
the sign of the polynomial $P(u)$, i.e.\ if we assume that the ``original'' wave oscillates, say, between $u_2$ and 
$u_3$, the oscillations in the conjugate equation occur between $u_1$ and $u_2$ with the (conjugate) 
wavenumber $\tilde k$.  It is not difficult to see that $\tilde k = 0$ at the harmonic edge and $\tilde k = O(1)$ 
at the solitary wave edge, so, indeed, $\tilde k$ behaves in a similar fashion to the amplitude.  As shown in 
\cite{el2,el1} the characteristic equation for the solitary wave edge has the same form (\ref{e:linbobore}), but with 
$k$ and $\omega$ replaced by their conjugate counterparts, and with the edge speed determined by the ratio 
$\tilde \omega / \tilde k$. 

As we have already stressed, the above construction is not applicable to the gBO equation (\ref{e:gbo}) due to the
qualitatively different structure of the equation for the travelling wave.  It turns out, however, that the solitary
wave edge characteristic in that case can be determined by a simple symmetry consideration, suggesting the correct 
choice of conjugate variables.  

In the following we shall assume that $V(u)$ is an odd function (which is the case for the standard 
BO equation \eqref{e:bo}). We now consider the mapping of DSW solutions of the initial value problem \eqref{e:gbo}  
and \eqref{e:boic} under the dispersion sign inversion $\sigma \to - \sigma$.  We shall call the gBO equations with $\sigma=-1$ 
and $\sigma=1$ the gBO$^-$ and the gBO$^+$ equations, respectively.  Note that, unlike the case of KdV type equations, the 
dispersion sign inversion cannot be achieved via the above mentioned complex change of independent variables $x \to i x$ and 
$t \to it$. However, another transformation $x \to -x$ and $u \to -u$ does map BO$^-$ to BO$^+$ (provided $V(u)$ 
is odd). This latter transformation implies a change of the DSW orientation and polarity, see \eqref{e:pol_or}.  
On the other hand, one can observe that the gBO {\it modulation system} (obtained, say, by applying the averaged Lagrangian 
formulation as in \cite{matsuno1, matsuno2} or by using multiple scales expansions as in \cite{ablowitz_2017}) is invariant 
with respect to the inversion of the dispersion sign, implying that the {\it modulation} solution of the gBO step problem 
maps to itself.  Indeed, in terms of modulation variables $\beta$, $k$ and $\omega$, where $\beta$ is the period averaged value 
of $u$, the dispersion sign inversion transformation $\sigma \to - \sigma$ corresponds to $\beta \to - \beta$ and $k \to - k$. 
Using the explicit expression for the averaged Lagrangian $\overline L$ for the BO$^-$ equation (see e.g.\ \cite{matsuno1}) one 
can see that this transformation implies $\overline L \to - \overline L$, so that the modulation equations are retained intact. 
Thus, in both cases one obtains the same system \eqref{e:modulation1} of decoupled simple wave equations. This symmetry 
of the averaged Lagrangian clearly remains the case for the gBO equation as well (although in the latter case the existence of
Riemann invariants is not guaranteed).  Combining the invariance of the modulation equations with the already established change  
of DSW orientation (see \eqref{e:pol_or}),  we conclude that, for the given initial data \eqref{e:boic}, the transformation 
$\sigma \to -\sigma$ maps the soliton edge of the DSW of the gBO$^-$ equation to the harmonic edge of the DSW of the gBO$^+$ 
equation, the latter being readily found using the approach of Sec.~\ref{sec:BO_harm}.

To this end  we assume for  definiteness that $\sigma =-1$  and use the transformation $\sigma \to - \sigma$ 
in \eqref{bo_disp} to obtain the conjugate dispersion relation $\tilde \omega = \tilde k V(\bar u) - \sigma \tilde k^2$  
corresponding to the gBO$^+$equation,  $\tilde k$ and $\tilde \omega$ being the wavenumber and frequency 
of the gBO$^+$ travelling  wave solution.  This immediately leads to the characteristic equation for the gBO$^+$ DSW harmonic 
edge (cf.\ (\ref{e:linbobore2}))
\begin{equation}
\frac{d \tilde k}{d\bar{u}} = \frac{1}{2\sigma} V'(\bar u).
\label{e:linbobore_conj}
\end{equation}
Since we have assumed that $\sigma=-1$, we have $d=-1$ for the conjugate DSW orientation (see \eqref{e:pol_or}) 
and so the solitary wave edge will be the trailing edge.  Then, the initial condition for 
(\ref{e:linbobore_conj}) is $\tilde k (u_-)=0$.  The solution of (\ref{e:linbobore_conj}) with $\sigma = -1$ is
\begin{equation}
\tilde k(\bar u) = \frac{1}{2} [V(u_-) - V({\bar u})],
\label{e:linwavenh}
\end{equation}
which gives the wavenumber at the harmonic (leading) edge of the ``conjugate'' DSW as
\begin{equation}
\tilde k_+= \tilde k(u_+) =  \frac{1}{2} [V(u_-) - V(u_{+})].
\end{equation}
Finally, the velocity $\tilde s_+$ of the leading, harmonic edge the conjugate DSW is determined in terms 
of the group velocity as
\begin{equation}
\tilde s_+ = \frac{\partial \tilde \omega}{\partial \tilde k} (\tilde k_+, u_+) = V(u_+) + 2 \tilde k_+ = V(u_{-}).
\label{e:grouplinbobnew}
\end{equation}
Equating $s_+ = \tilde s_+$, we obtain the sought solitary wave edge velocity of the original BO$^-$ DSW.
In particular, for the case of the integrable BO equation (\ref{e:bo}) with $V(u)=2u$ and $\sigma=-1$ we 
obtain $s_+= 2u_-$, which agrees with the result of \cite{tim,matsuno1} reproduced in Sec.\ \ref{sec:int_theory}. 
The lead soliton amplitude $a_s=4 (u_- - u_+)$ then follows from equating $s_+=c_s$ in (\ref{e:bosol}) and 
assuming the background $\bar u=u_+$.

\section{Riemann problem for Calogero-Sutherland dispersive hydrodynamics}

Now that the DSW fitting method \cite{el2,el1} for the determination of the leading
and trailing edges of DSWs has been extended to equations with BO type dispersion and verified for the case of 
the standard BO equation (\ref{e:bo}), it will now be applied to the CS system (\ref{e:mass1}) and (\ref{e:mom1}) 
for which there is no known DSW solution.  Although the CS system is completely integrable \cite{abanov2} and its 
modulation system in Riemann form can, in principle, be obtained, the DSW fitting approach is more direct and also 
offers the important advantage of generalisation to non-integrable equations (see the discussion in the 
Introduction). 
%\begin{eqnarray}
% &&\frac{\partial \rho}{\partial t} + \frac{\partial}{\partial x} \rho v  =  0, \label{e:mass} \\
%&&  \frac{\partial v}{\partial t} + \frac{\partial}{\partial x} \left[ \frac{1}{2}v^{2} + \frac{1}{2}
%\pi^{2} g^{2} \rho^{2} -  g^2 \left( \frac{1}{4} \frac{\rho_{xx}}{\rho} + \frac{1}{8} \frac{\rho_{x}^{2}}
%{\rho^{2}}  + \pi  \mathcal{H}\rho_{x}  \right)\right]  =  0 , \label{e:mom}
%\end{eqnarray}
%where again $\mathcal{H}$ denotes the Hilbert transform.  

\begin{figure}[h]
\centering
\includegraphics[width=0.6\textwidth]{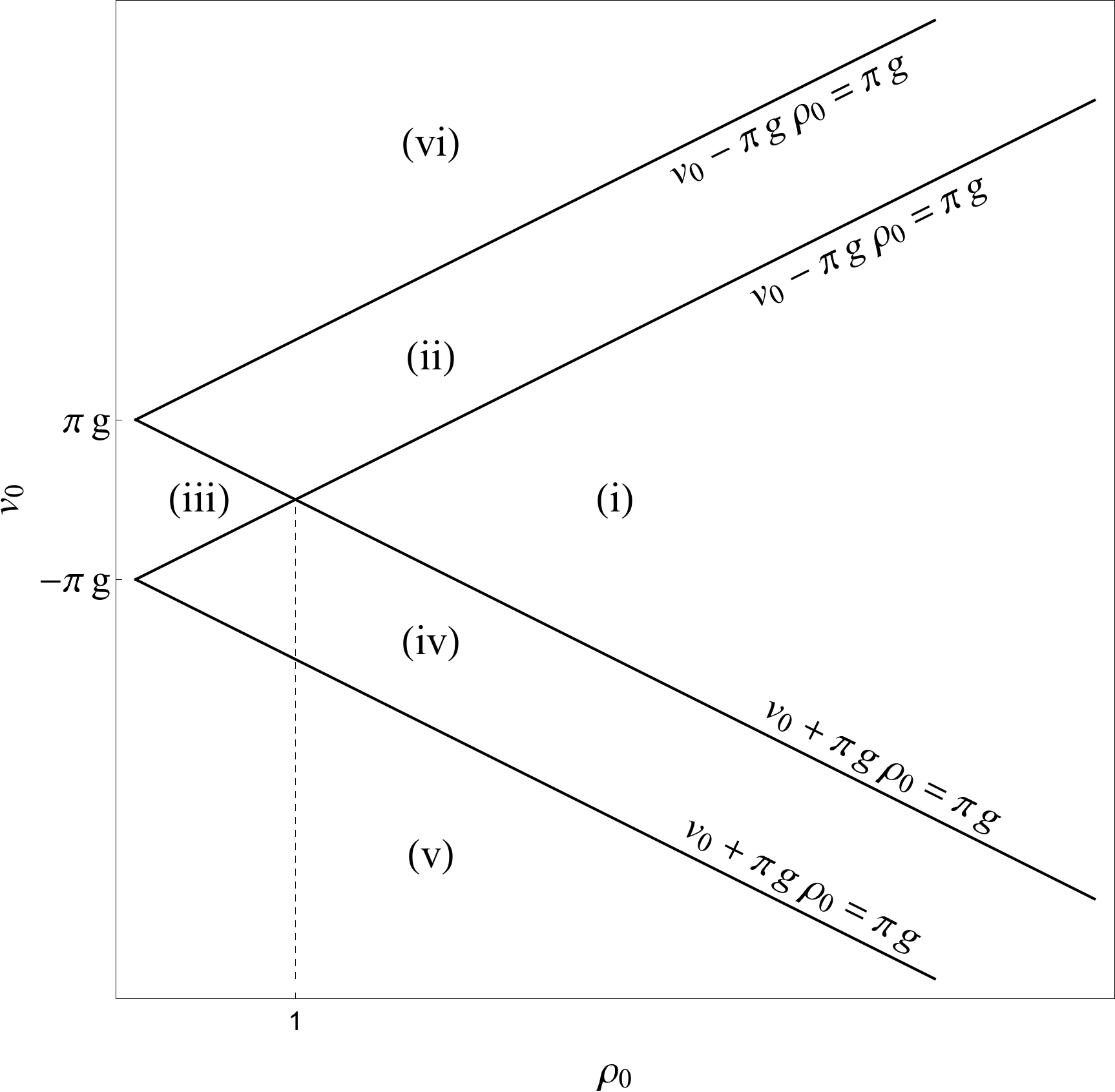}
\caption{Riemann problem classification for the CS system (\ref{e:mass1}) and (\ref{e:mom1}).  The initial step 
(\ref{e:rhoic}) parameters are normalised to $\rho_+=1$, $v_+=0$, $\rho_-=\rho_0$ and $v_-=v_0$.}
\label{f:class1}
\end{figure}

\begin{figure}
\centering
\includegraphics[width=0.33\textwidth,angle=270]{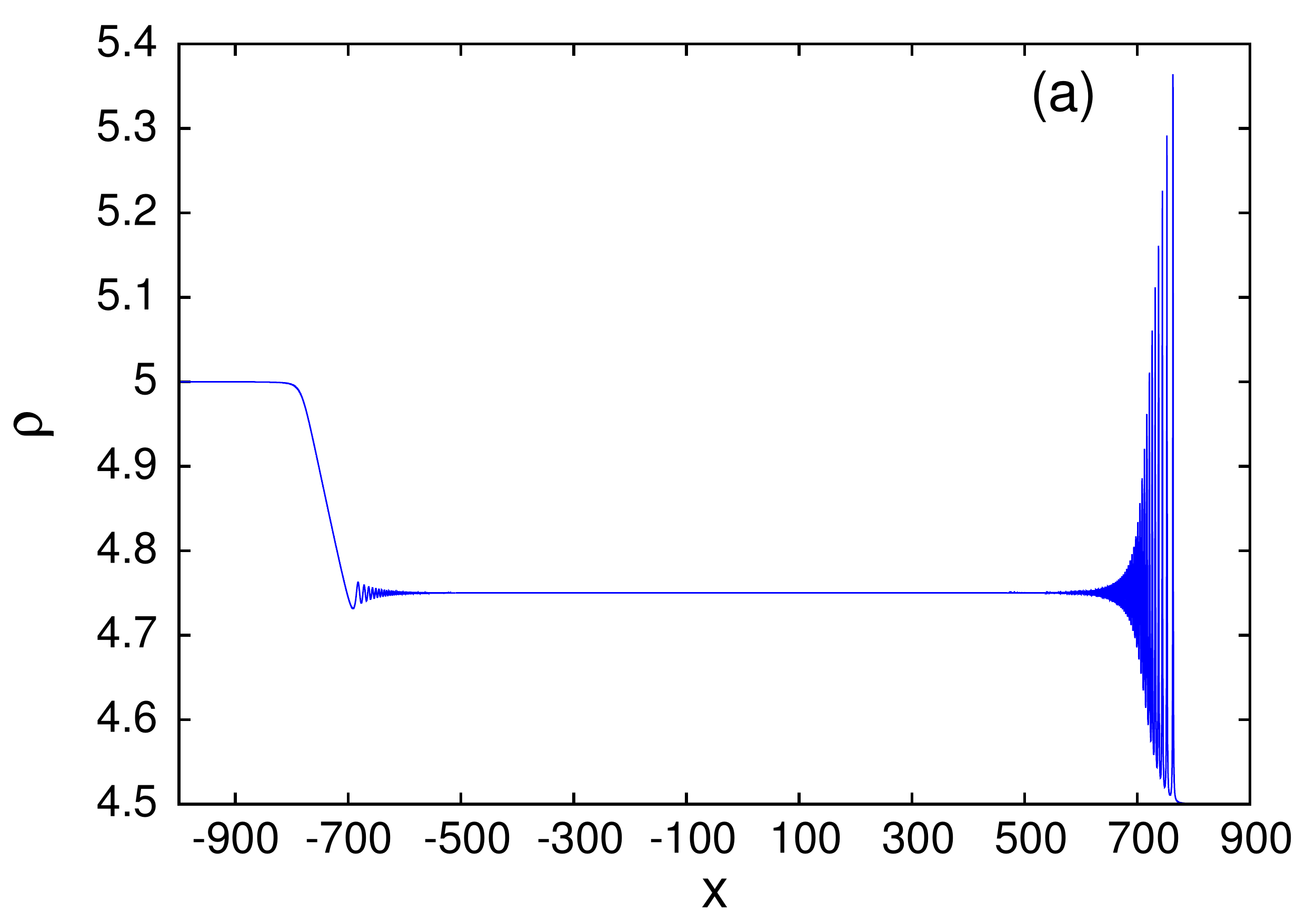}
\includegraphics[width=0.33\textwidth,angle=270]{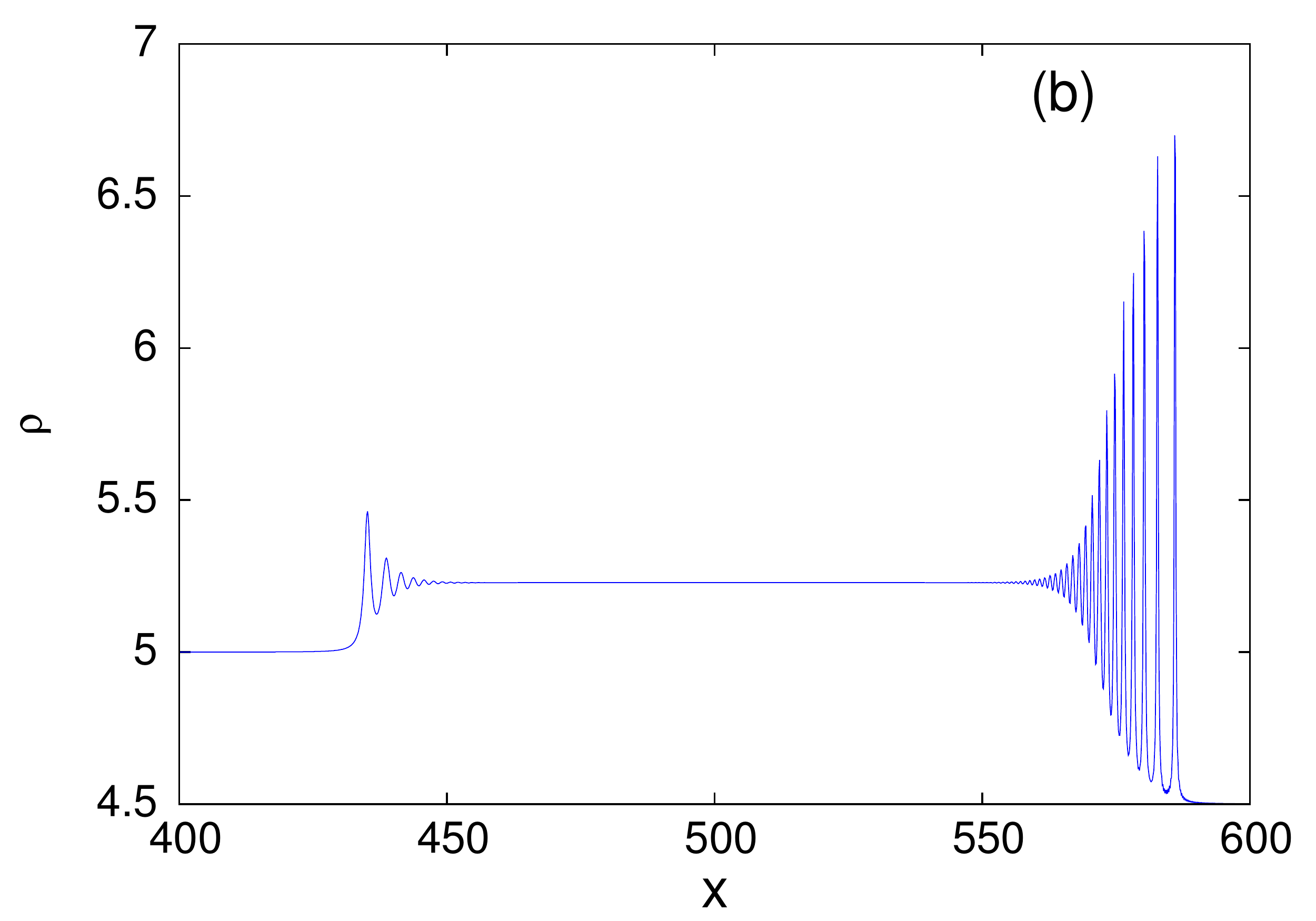}
\includegraphics[width=0.33\textwidth,angle=270]{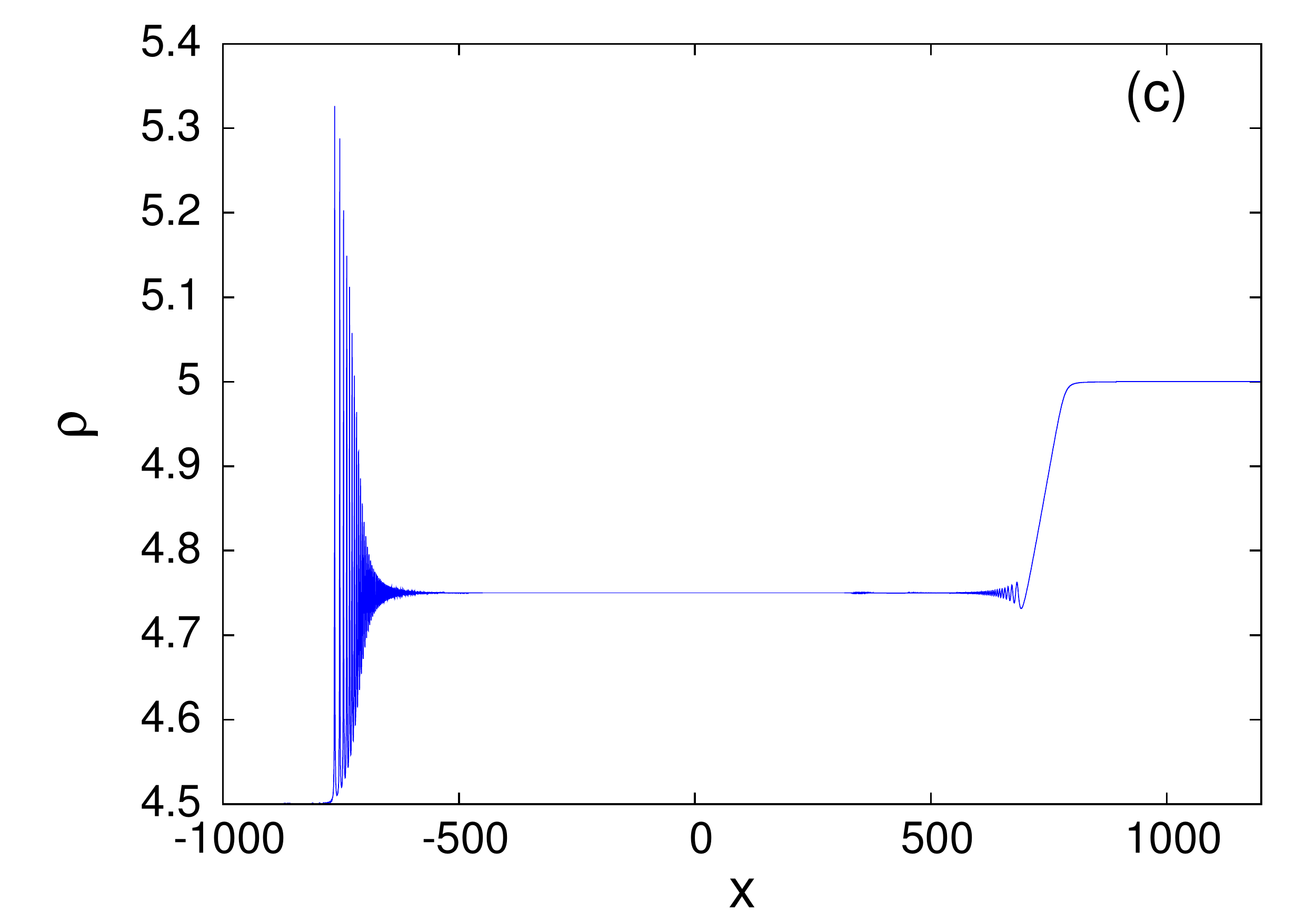}
\includegraphics[width=0.33\textwidth,angle=270]{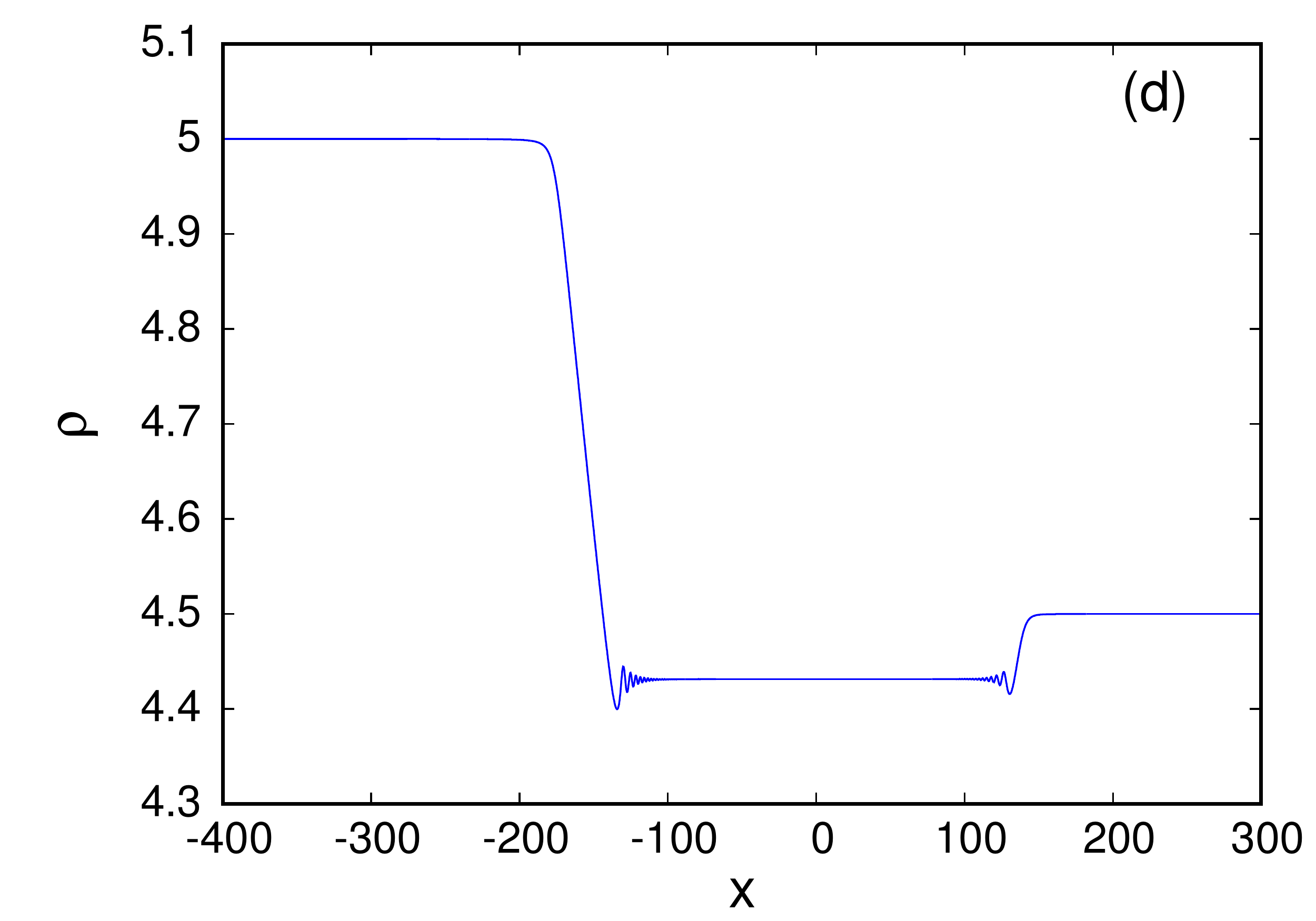}
\includegraphics[width=0.33\textwidth,angle=270]{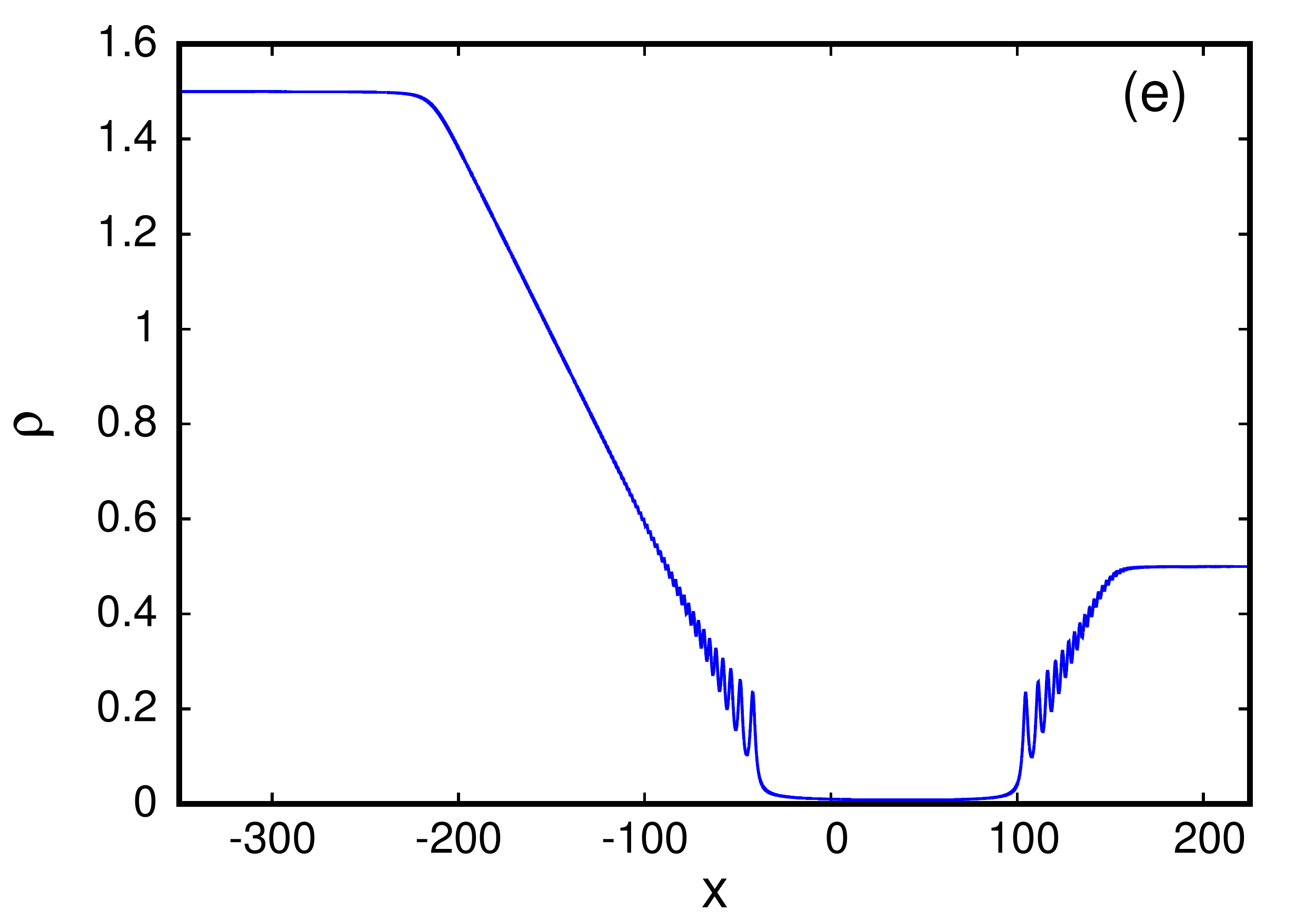}
\includegraphics[width=0.33\textwidth,angle=270]{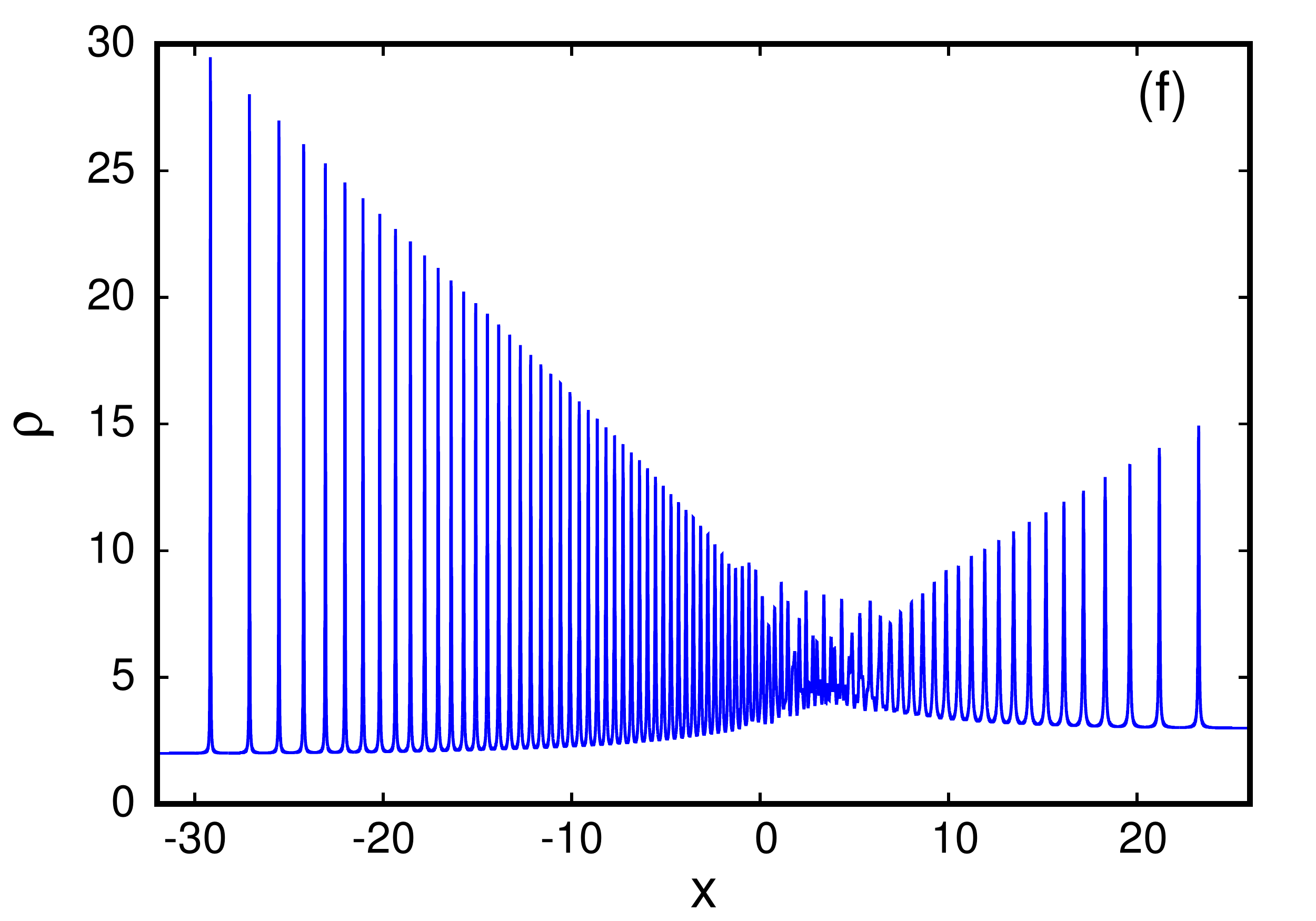}
\caption{Examples of the six solution forms for the CS Riemann problem. Solution for $\rho$. (a) Case (i), $\rho_{+}=4.5$, 
$\rho_{-}=5$, $v_{+}=v_{-}=0$, $t=100$; (b) Case (ii) $\rho_{+}=4.5$, $\rho_{-}=5$, $v_{+}=0$, $v_{-}=1.5$, $t=10$; 
(c) Case (iii), $\rho_{+}=5$, $\rho_{-}=4.5$, $v_{+}=v_{-}=0$, $t=100$; (d) Case (iv) $\rho_{+}=4.5$, $\rho_{-}=5$, 
$v_{+}=0$, $v_{-}=-1.0$, $t=20$; (e) Case (v) $\rho_{+} = 0.5$, $\rho_{-}=1.5$, $v_{+}=3$, $v_{-}=-3$, $t=40$;
(f) Case (vi) $\rho_{+} = 3$, $\rho_{-}=2$, $v_{+}=-4$, $v_{-}=4$, $t=4$.  $g = 0.5$.}
\label{f:casesrho}
\end{figure}

\begin{figure}
\centering
\includegraphics[width=0.33\textwidth,angle=270]{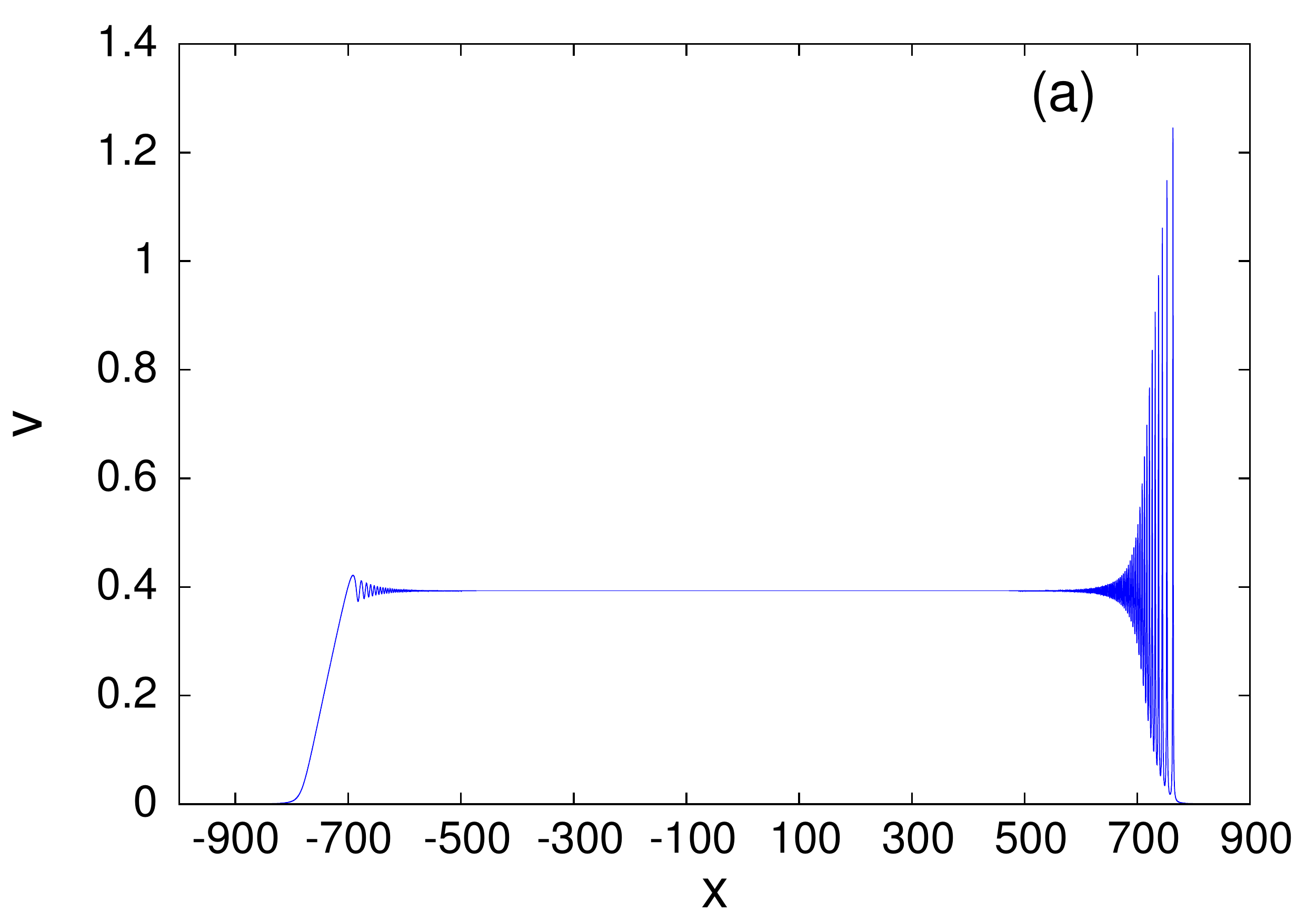}
\includegraphics[width=0.33\textwidth,angle=270]{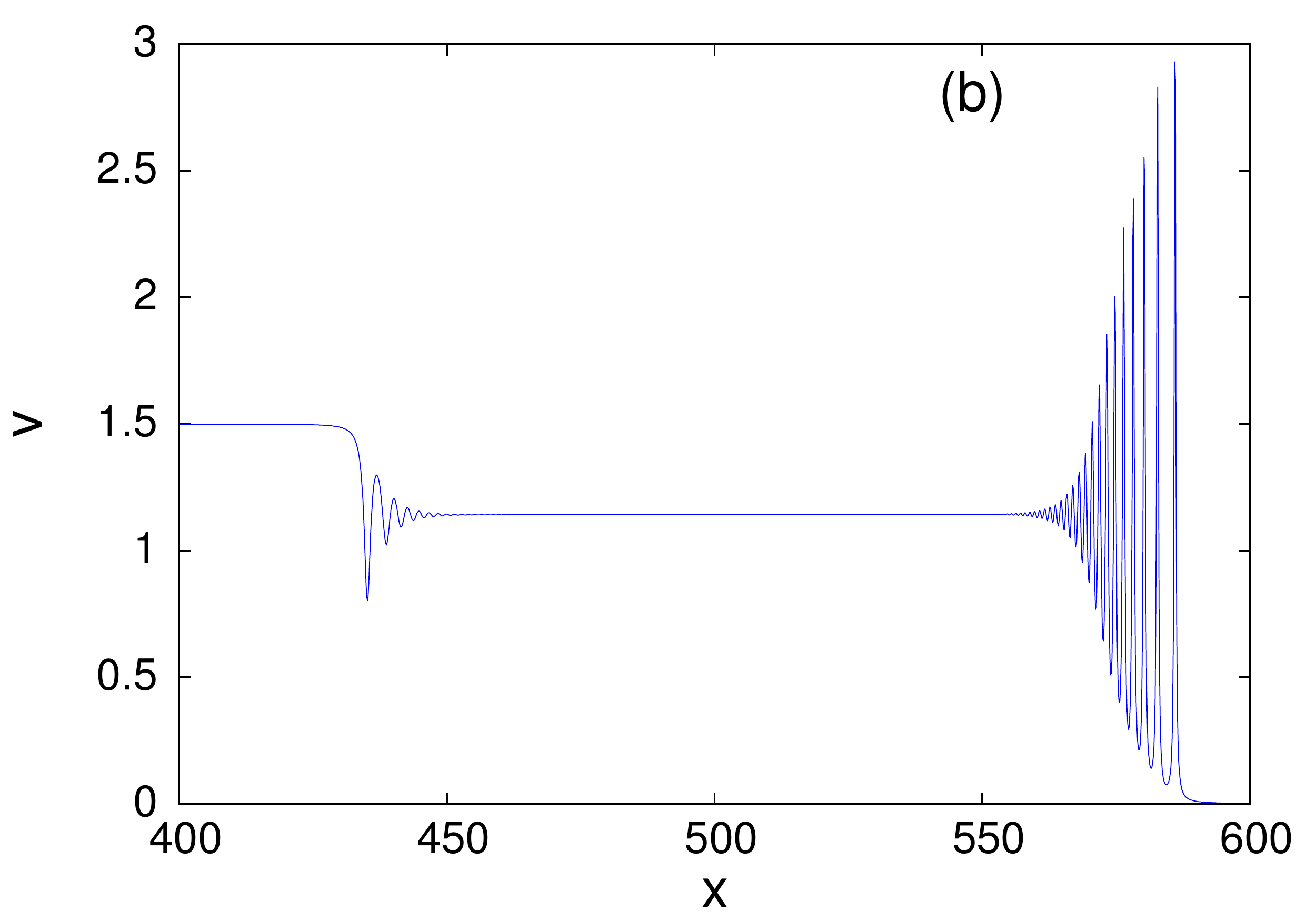}
\includegraphics[width=0.33\textwidth,angle=270]{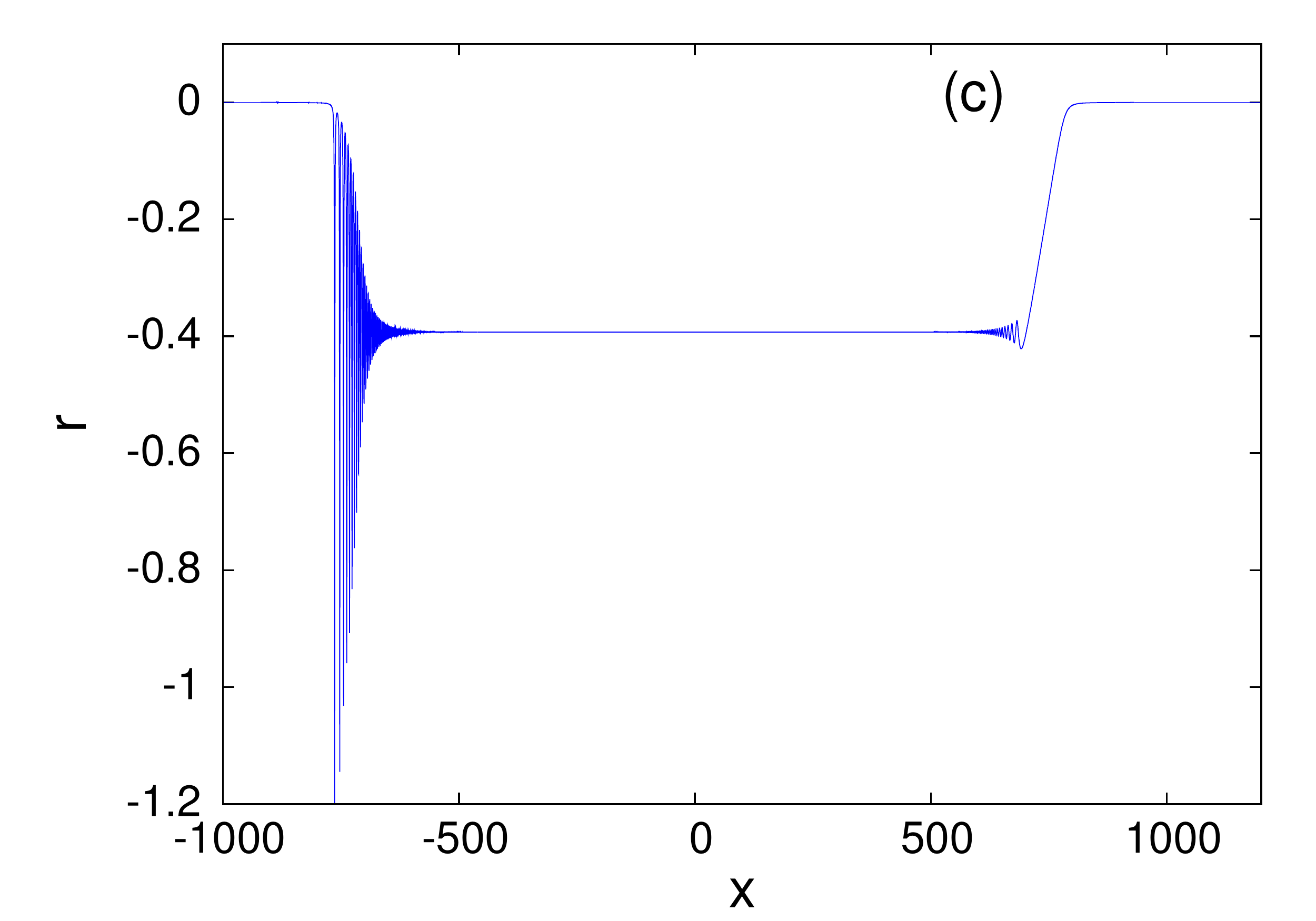}
\includegraphics[width=0.33\textwidth,angle=270]{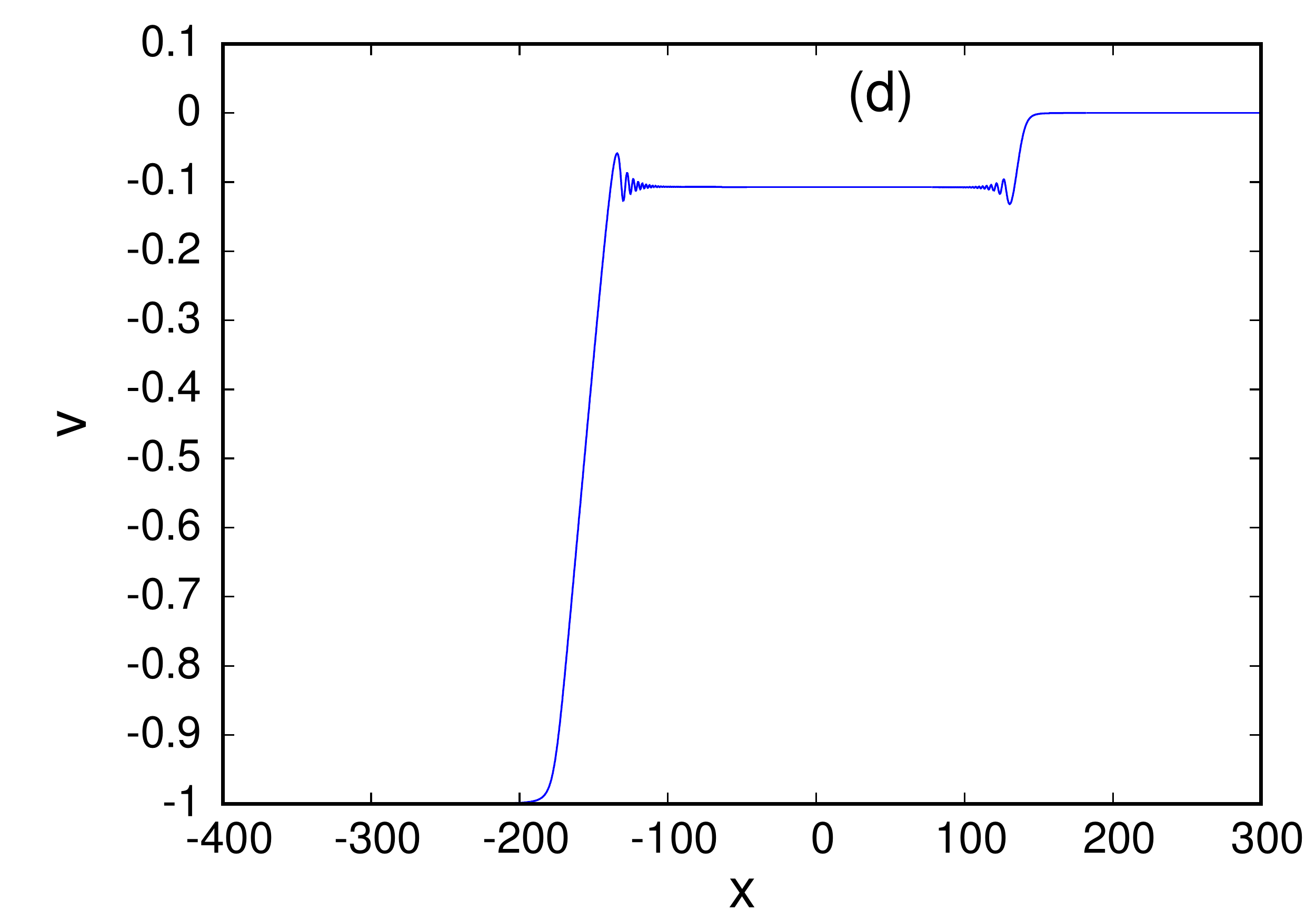}
\includegraphics[width=0.33\textwidth,angle=270]{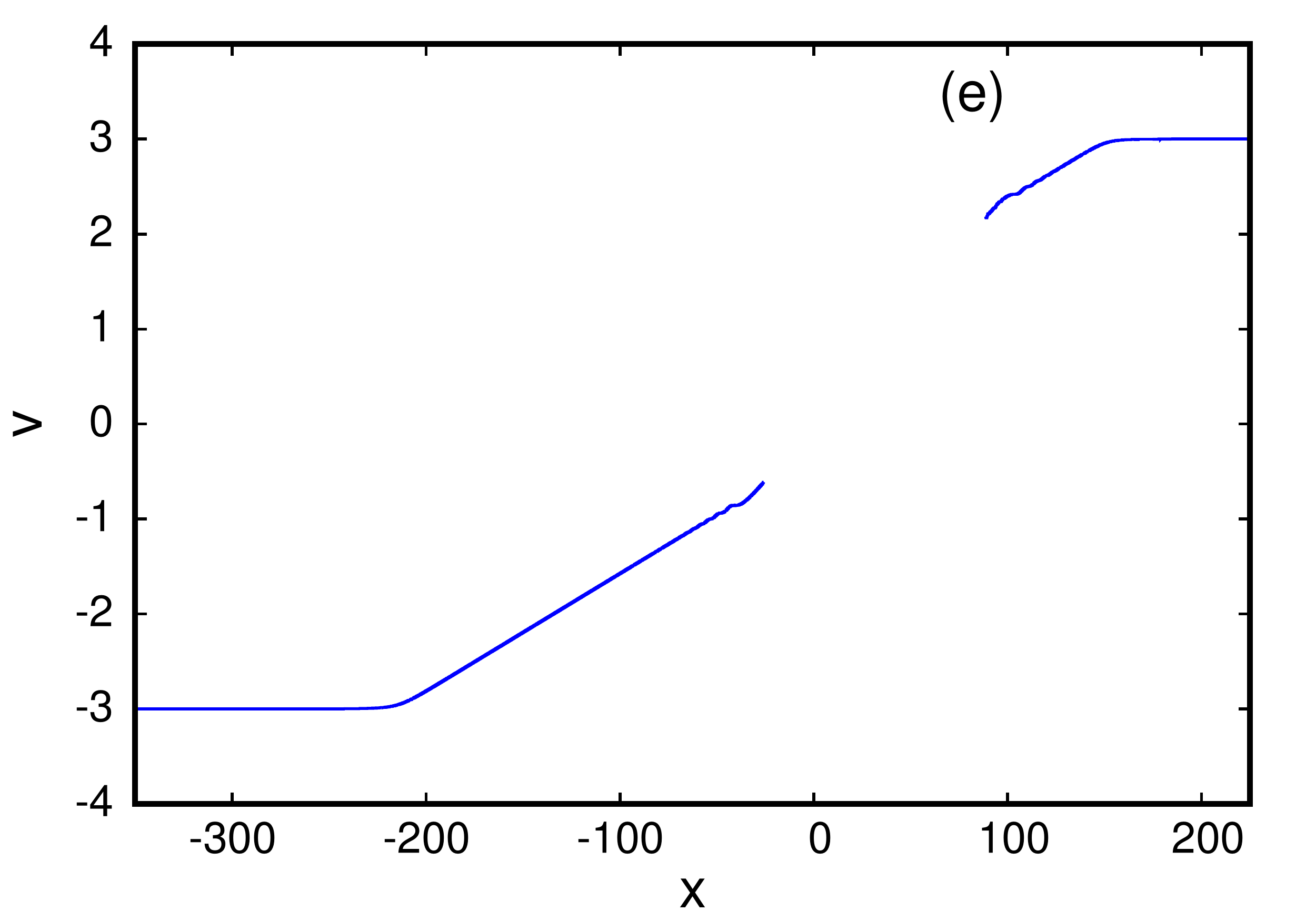}
\includegraphics[width=0.33\textwidth,angle=270]{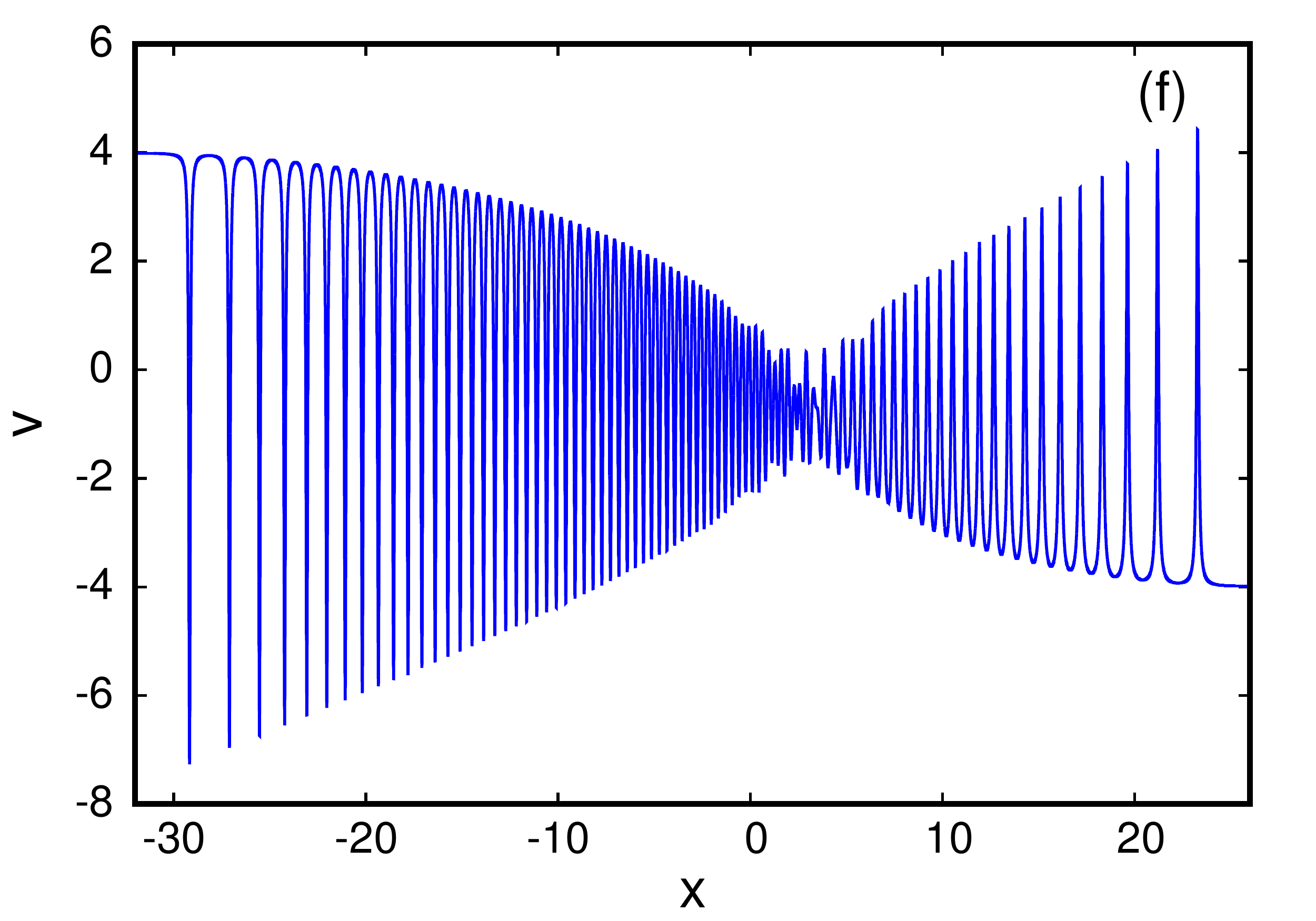}
\caption{Examples of the six solution forms for the CS Riemann problem. Solution for $v$. (a) Case (i), $\rho_{+}=4.5$, 
$\rho_{-}=5$, $v_{+}=v_{-}=0$, $t=100$; (b) Case (ii) $\rho_{+}=4.5$, $\rho_{-}=5$, $v_{+}=0$, $v_{-}=1.5$, $t=10$; 
(c) Case (iii), $\rho_{+}=5$, $\rho_{-}=4.5$, $v_{+}=v_{-}=0$, $t=100$; (d) Case (iv) $\rho_{+}=4.5$, $\rho_{-}=5$, 
$v_{+}=0$, $v_{-}=-1.0$, $t=20$; (e) Case (v) $\rho_{+} = 0.5$, $\rho_{-}=1.5$, $v_{+}=3$, $v_{-}=-3$, $t=40$;
(f) Case (vi) $\rho_{+} = 3$, $\rho_{-}=2$, $v_{+}=-4$, $v_{-}=4$, $t=4$.  $g = 0.5$.}
\label{f:casesv}
\end{figure}

The CS equations (\ref{e:mass1}) and (\ref{e:mom1}) will be solved with the step initial conditions
\begin{equation}
 \rho = \left\{ \begin{array}{cc}
                 \rho_{-}, & x < 0, \\
                 \rho_{+}, & x > 0
                \end{array}
        \right. , \qquad
  v = \left\{ \begin{array}{cc}
                 v_{-}, & x < 0, \\
                 v_{+}, & x > 0.
                \end{array}
        \right.       
\label{e:rhoic}
\end{equation}
We shall assume that $\rho_->0$ and $\rho_+>0$. The special case $\rho_+=0$, the dam break problem \cite{whitham}, 
will be considered separately.
%and
%\begin{equation}
% v = \left\{ \begin{array}{cc}
%                 v_{-}, & x < 0, \\
%                 v_{+}, & x > 0.
%                \end{array}
%        \right. 
%\label{e:vic}
%\end{equation}
%Away from a DSW dispersion plays a negligible role in the solutions of the CS equations 
%(\ref{e:mass1}) and (\ref{e:mom1}) \cite{borereview}.  
 The modulation equations for the CS system (\ref{e:mass1}) and (\ref{e:mom1}) have the same type 
of  symmetry under the reflection $x \to -x$ and $v \to -v$, accompanied by the change of the sign of the nonlocal 
dispersion,  as the BO equation, which suggests the amenability of the modification of the DSW fitting approach 
developed in Section \ref{s:bo} to the CS system.  However, as the CS system is bi-directional the solution of 
a Riemann problem is more complex than that for the BO equation, consisting of various combinations of DSWs and 
expansion fans, as for the defocusing NLS equation \cite{elnls}.  Our analysis below shows that, as for the 
defocusing NLS Riemann problem solution \cite{elnls}, there are six possible solutions of the CS equations with 
the initial condition (\ref{e:rhoic}). The classification diagram for normalised Riemann data (\ref{e:rhoic}) 
with $\rho_+=1$ and $v_+=0$ is presented in Figure \ref{f:class1}.  The solution forms corresponding to different 
regions of the diagram in Fig.\ \ref{f:class1} are illustrated in Figs.\ \ref{f:casesrho} and \ref{f:casesv} 
and are:

\begin{enumerate}

 \item  A DSW propagating downstream, followed by an expansion wave propagating upstream, separated by a steady plateau 
 (Figs.\ \ref{f:casesrho}(a), \ref{f:casesv}(a)).
 
 \item  Two DSWs, propagating upstream and downstream, separated by a steady plateau (Figs.\ \ref{f:casesrho}(b), 
 \ref{f:casesv}(b)).
 
 \item  A DSW propagating upstream, followed by an expansion wave propagating downstream, separated by a steady plateau 
 (Figs.\ \ref{f:casesrho}(c), \ref{f:casesv}(c)).
 
 \item  Two expansion waves separated by a constant plateau (Figs.\ \ref{f:casesrho}(d), \ref{f:casesv}(d)).
 
 \item  Two expansion waves separated by a vacuum region with $\rho=0$ (Figs.\ \ref{f:casesrho}(e), \ref{f:casesv}(e)).
        Note that the phase $v$ is undefined in the vacuum region $\rho=0$.
 
 \item  Upstream and downstream overlapping DSWs (Figs.\ \ref{f:casesrho}(f), \ref{f:casesv}(f)).
 
\end{enumerate}

%\begin{figure}
%\centering
%\includegraphics[width=0.33\textwidth,angle=270]{case1rho.eps}
%\includegraphics[width=0.33\textwidth,angle=270]{case2rho.eps}
%\includegraphics[width=0.33\textwidth,angle=270]{case3rho.eps}
%\includegraphics[width=0.33\textwidth,angle=270]{case4rho.eps}
%\caption{Examples of six solution forms for CS Riemann problem. Solution for $\rho$. (a) Case (i), $\rho_{+}=4.5$, 
%$\rho_{-}=5$, $v_{+}=v_{-}=0$, $t=100$; (b) Case (ii) $\rho_{+}=4.5$, $\rho_{-}=5$, $v_{+}=0$, $v_{-}=1.5$, $t=10$; 
%(c) Case (iii), $\rho_{+}=5$, $\rho_{-}=4.5$, $v_{+}=v_{-}=0$, $t=100$; (d) Case (iv) $\rho_{+}=4.5$, $\rho_{-}=5$, 
%$v_{+}=0$, $v_{-}=-1.0$, $t=20$.  $g = 0.5$.}
%\label{f:casesrho}
%\end{figure}

The key ingredient of the DSW fitting method is the linear dispersion relation, which has the form (\ref{dr_2BO}) 
for the CS system.  Due to bi-directionality, the solution of the full Riemann problem will also involve the simple 
wave solutions of the dispersionless limit of the CS system, which has the form
\begin{eqnarray}
 \frac{\partial \rho}{\partial t} + v\frac{\partial \rho}{\partial x} + \rho \frac{\partial v}{\partial x} 
& = & 0, \label{e:massless} \\
 \frac{\partial v}{\partial t} + v\frac{\partial v}{\partial x} + \pi^{2}g^{2} \rho \frac{\partial \rho}{\partial x} 
& = & 0 . \label{e:momles}
\end{eqnarray}
The system (\ref{e:massless}) and (\ref{e:momles}) represents a system of gas dynamics equations for an isentropic 
polytropic gas with the pressure law $P(\rho) = \frac 13 \pi^2 g^2 \rho^3$ \cite{whitham}.  It can be readily 
represented in the Riemann invariant form 
\begin{equation}
\frac{\partial r_{\pm}}{\partial t} + V_\pm(r_+, r_-) \frac{\partial r_\pm}{\partial x} = 0, \quad i=1,2, 
\label{e:riemann}
\end{equation}
with the Riemann invariants $r_{\pm}$ and the characteristic velocities $V_{\pm}$ found as \cite{whitham} 
\begin{equation}\label{e:v12}
V_\pm = v \pm c(\rho), \quad  r_{\pm} = v \pm \int \frac{c(\rho)}{\rho} d\rho.
\end{equation}
Here, $c(\rho)= \sqrt{dP/d\rho}$ is the long wave speed of sound.  In the CS fluid case $c(\rho) = \pi g \rho$ and the 
Riemann invariant form of the dispersionless system (\ref{e:massless}) and (\ref{e:momles}) becomes a system of two 
uncoupled Hopf equations
\begin{equation}\label{e:dless}
\frac{\partial r_+}{\partial t}  + r_+\frac{\partial r_+}{\partial x}  =0, \quad  \frac{\partial r_-}{\partial t}  + 
r_-\frac{\partial r_-}{\partial x}  =0,
\end{equation}
where 
\begin{equation}\label{e:dless_r}
r_\pm=  v \pm \pi g \rho.
\end{equation}
%\begin{eqnarray}
% C_{+}: \quad \frac{d}{dt} \left( \pi g \rho + v \right) = 0 & \quad \mbox{on} \quad & \frac{dx}{dt} = v + \pi g \rho = V_{+}, 
% \label{e:vp} \\
% C_{-}: \quad \frac{d}{dt} \left( \pi g \rho - v \right) = 0 & \quad \mbox{on} \quad & \frac{dx}{dt} = v - \pi g \rho = V_{-} . 
% \label{e:vm}
%\end{eqnarray}
We shall call the characteristics $dx/dt=r_+$ ($dx/dt=r_-$) the $C_{+}$  ($C_-$) characteristics.  We note 
that the above gas dynamic interpretation of the dispersionless equations (\ref{e:massless}) and (\ref{e:momles}) 
offers a natural generalisation of the CS system (\ref{e:mass1}) and (\ref{e:mom1}) by considering a different 
pressure law, e.g.\ $P(\rho) \sim \rho^2$, which would constitute an NLS-BO system (presumably non-integrable), 
which could also be considered by the method developed here.

%Simple wave solutions of this dispersionless system will be used below in constructing solutions of the 
%CS equations for the initial conditions (\ref{e:rhoic}).  

%To obtain the linear dispersion relation we linearise the system (\ref{e:mass}) and (\ref{e:mom}) with
%\begin{equation}
% \rho = \bar{\rho} + \rho_{1}, \quad v = \bar{v} + v_{1},
%\label{e:linvar}
%\end{equation}
%and $|\rho_{1}| \ll \bar{\rho}$ and $|v_{1}| \ll |\bar{v}|$.  The linearised system is then
%\begin{eqnarray}
% \frac{\partial \rho_{1}}{\partial t} + \bar{v}\frac{\partial \rho_{1}}{\partial x} + 
% \bar{\rho} \frac{\partial v_{1}}{\partial x} & = & 0, \label{e:masslin} \\
%\frac{\partial v_{1}}{\partial t} + \bar{v} \frac{\partial v_{1}}{\partial x} + \pi^{2}g^{2}\bar{\rho}
%\frac{\partial \rho_{1}}{\partial x} - \frac{g^{2}}{4\bar{\rho}} \frac{\partial^{3}\rho_{1}}{\partial x^{3}} 
%+ \pi g^{2}H\rho_{1xx} & = & 0 . \label{e:momlin} 
%\end{eqnarray}
%We now look for travelling wave solutions
%\begin{equation}
%\rho_{1} = Ae^{i(kx - \omega t)}, \quad v_{1} = Be^{i(kx - \omega t)},
%\label{e:travel} 
%\end{equation}
%from which the dispersion relation can be found to be
%\begin{equation}
% \omega = k\bar{v} \pm \frac{1}{2}gk(k - 2\pi \bar{\rho}) .
%\label{e:drel}
%\end{equation}
%

%The Calogero-Sutherland equations (\ref{e:mass}) and (\ref{e:mom}) are bi-directional, so that its 
%DSW solutions bear more similarity to those for the equivalent NLS equation \cite{elnls} than the Benjamin-Ono
%equation of the previous section
We shall now find the solutions for the first five cases introduced at the beginning of this section by
using combinations of simple wave solutions of the dispersionless equations (\ref{e:dless}) and (\ref{e:dless_r}) 
and the relations describing the DSW closure as determined by the DSW fitting method.  Unfortunately, the solution 
for case (vi), consisting of two partially realised DSWs connected by an oscillating shelf, requires the full Whitham 
modulation equations, see \cite{borereview,elnls} for the description of a similar case for the defocusing NLS equation.
This case will then be investigated numerically only.  In Section \ref{s:comp} the resulting modulation theory solutions 
will be compared with full numerical solutions of the CS system (\ref{e:mass1}) and (\ref{e:mom1}).

\subsection{Case (i):  fast DSW and slow expansion wave}
\label{s:case1}

The solution for this case consists of a slow (associated with the $C_-$ characteristic) expansion wave linking the 
levels $\rho = \rho_{-}$ and $v = v_{-}$ to an intermediate shelf $\rho = \rho_{i}$ and $v = v_{i}$.  The solution 
is then taken from this intermediate shelf $\rho = \rho_{i}$ and $v = v_{i}$ to the levels $\rho = \rho_{+}$ and 
$v = v_{+}$ by a fast DSW. The solution form for this case is illustrated in Figures \ref{f:casesrho}(a) and \ref{f:casesv}(a).  

The intermediate level $\rho = \rho_{i}$ and $v = v_{i}$ between the expansion wave and the DSW can be determined 
from the dispersionless system (\ref{e:dless}) and (\ref{e:dless_r}).  The slow expansion wave has the Riemann 
invariant $r_+$ constant through it.  Hence,
\begin{equation}
 v_{i} + \pi g \rho_{i}  =   v_{-} +  \pi g \rho_{-}.
\label{e:cp}
\end{equation}
In addition, as was proposed in \cite{gur_meshch}, and proved using modulation theory in \cite{el2} (see also 
\cite{borereview}) the fast DSW locus is determined by the conservation of the dispersionless Riemann invariant 
$r_-$ (\ref{e:dless_r}) across the DSW, so that
\begin{equation}
v_i -  \pi g \rho_{i}  = v_+ - \pi g \rho_{+}.
\label{e:cm}
\end{equation}
Note that condition (\ref{e:cm}), unlike in the expansion fan case (\ref{e:cp}), does not imply the conservation of 
$r_-$ {\it within} the DSW.  Eliminating between (\ref{e:cp}) and (\ref{e:cm}), we obtain the intermediate level
\begin{equation}
 \rho_{i} = \frac{1}{2} \left[ \rho_{+} + \rho_{-} + \frac{v_{-} - v_{+}}{\pi g} \right], \quad
v_{i} = \frac{\pi g}{2} \left[ \rho_{-} - \rho_{+} + \frac{v_{-} + v_{+}}{\pi g} \right].
\label{e:inter}
\end{equation}
With this intermediate level, the expansion fan from the level $\rho_{-}$ to $\rho_{i}$ can now be determined.

The expansion fan linking the level $\rho_{-}$ to the intermediate shelf $\rho = \rho_{i}$ is a similarity simple wave 
solution with $r_+=const$ on the $C_-$ characteristics of the dispersionless system (\ref{e:massless}) and (\ref{e:momles}).  
It is then found on using the intermediate shelf solution (\ref{e:inter}) that this expansion fan weak solution is
\begin{equation}
 \rho = \left\{ \begin{array}{cc}
                \rho_{-}, & \frac{x}{t} < v_{-} - \pi g \rho_{-}, \\
                \frac{\pi g\rho_{-} + v_{-} - \frac{x}{t}}{2\pi g}, & v_{-} - \pi g \rho_{-} \le \frac{x}{t} \le
                v_{+} - \pi g \rho_{+}, \\
                \rho_{i}, & v_{+} - \pi g \rho_{+} < \frac{x}{t} <  s_{i}
               \end{array}
        \right. 
\label{e:rhofan}
\end{equation}
and
\begin{equation}
 v = \left\{ \begin{array}{cc}
                v_{-}, & \frac{x}{t} < v_{-} - \pi g \rho_{-}, \\
                \frac{x}{2t} + \frac{1}{2}\pi g \rho_{-} + \frac{1}{2}v_{-}, & v_{-} - \pi g \rho_{-} \le \frac{x}{t} \le
                v_{+} - \pi g \rho_{+}, \\
                v_{i}, & v_{+} - \pi g \rho_{+} < \frac{x}{t} < s_{i}.
               \end{array}
        \right. 
\label{e:vfan}
\end{equation}
The right hand end of the intermediate shelf $(\rho_i, v_i)$ is terminated at the position $x = s_i t$, which is the position
of the trailing edge of the DSW. The expansion fan solution is realised if
\begin{equation}\label{e:fancond}
v_{-} - \pi g\rho_{-} < v_{+} - \pi g\rho_{+}.
\end{equation} 
With the expansion fan and the intermediate level now determined, the DSW linking the intermediate level to
the level ahead $(\rho_{+},v_{+})$ can be determined.  As for the BO DSW of Section \ref{s:bo} the leading and
trailing edges of this DSW are determined from the linear dispersion relation (\ref{dr_2BO}) of the 
CS system (\ref{e:mass1}) and (\ref{e:mom1}). This linear dispersion relation must be augmented by the simple wave relation 
$\bar v(\bar \rho)$ due to the DSW locus (\ref{e:cm}) \cite{el2,borereview,hoefer},
\begin{equation}
 \bar{v} = v_{+} - \pi g \rho_{+} + \pi g \bar{\rho} ,
\label{e:vbore}
\end{equation}
giving the frequency
\begin{equation}
\Omega(k, \bar \rho)= \omega (k, \bar \rho, \bar v (\bar \rho)) = \left( v_{+} - \pi g \rho_{+} \right) k + 2\pi g \bar{\rho}k 
- \frac{g}{2} k^{2},
\label{e:omegam}
\end{equation}
%where we have chosen the ``minus'' branch in the dispersion relation (\ref{dr_2BO}) corresponding to the upstream DSW. 
The characteristic relation $k(\bar \rho)$ for the trailing, harmonic edge of the DSW is then determined by the ODE
\begin{equation}
 \frac{dk}{d\bar{\rho}} = \frac{\frac{\partial \Omega}{\partial\bar{\rho}}}{V_{+} - \frac{\partial \Omega}{\partial k}}, 
\label{e:linode}
\end{equation}
where (see (\ref{e:v12}) and (\ref{e:vbore})) 
\begin{equation}
 V_{+} = V_+(\bar \rho, \bar v(\rho)) = v_{+} - \pi g \rho_{+} + 2\pi g \bar{\rho} .
\label{e:vpbore}
\end{equation}
The solution of the differential equation (\ref{e:linode}) with the boundary condition $k=0$ at $\bar{\rho} = \rho_{+}$,
ensuring consistency with the leading, solitary wave edge of the DSW (see \cite{el2,borereview,el1}), is
\begin{equation}
 k = 2\pi \left( \bar{\rho} - \rho_{+}\right).
\label{e:ksoln}
\end{equation}
We finally have that the wavenumber and velocity of the trailing edge of the downstream DSW are
\begin{equation}
 k_i=k(\rho_i) = 2\pi \left( \rho_{i} - \rho_{+} \right), \quad  s_i= \frac{\partial \Omega}{\partial k}(\rho_i, k_i) = 
v_{+} + \pi g \rho_{+} ,
\label{e:grouplintrail}
\end{equation}
respectively. 

As for the BO equation of Section \ref{s:bo}, the leading, solitary wave edge of the DSW can be found on using 
the dispersion sign transformation, so that the conjugate CS system (an analogue of the  BO$^+$ equation in 
Sec.\ \ref{sec:sol_edge_BO}) has the dispersion relation for right-going waves (cf.\ (\ref{dr_2BO}))
\begin{equation} \label{dr_2BO_conj}
\tilde \omega = \tilde kV_{+} (\bar{\rho}, \bar{v}) + \frac{1}{2}g \tilde k^2.
\end{equation}
One can see from (\ref{dr_2BO_conj}) that $\tilde \omega_{\tilde k \tilde k}>0$, so that the conjugate DSW in the 
$(-x, -t)$ plane has negative polarity and orientation, so that its harmonic edge coincides with the solitary 
wave edge of the original CS DSW (see Sec.~\ref{sec:sol_edge_BO} for the explanation of a similar mapping for 
the BO DSW).  

The locus of the conjugate DSW is given by Eq.\ (\ref{e:cm}), which is the same as for the original DSW.  The conjugate 
dispersion relation is augmented by Eq.\ (\ref{e:vbore}) relating $\bar v$ and $\bar \rho$, giving the conjugate frequency
\begin{equation}
\tilde \Omega (\tilde k, \bar \rho) = \tilde \omega (k, \bar \rho, \bar v (\bar \rho)) = \left( v_{+} - \pi g \rho_{+} \right) 
\tilde k + 2\pi g \bar{\rho}k - \frac{1}{2}g \tilde k^{2}.
\end{equation}
The characteristic ODE for $\tilde k (\bar \rho)$ at the harmonic (leading) edge of the conjugate DSW then has the same 
form (\ref{e:linode}), but with $k$ replaced by $\tilde k$ and $\Omega$ by $\tilde \Omega$.  It is solved with 
the boundary condition $\tilde k (\rho_i)=0$. The solution is then readily found to be
\begin{equation}
\tilde k = 2\pi(\rho_i -\bar \rho).
\end{equation}
The velocity of the harmonic edge of the conjugate DSW is finally
\begin{equation}
\tilde s_+ = \frac{\partial \tilde \Omega}{\partial \tilde k} (\tilde k(\rho_+), \rho_+) = v_+-\pi g \rho_+ +
2 \pi g \rho_i \, .
\end{equation}
Using (\ref{e:cp}) and (\ref{e:cm}) we obtain 
\begin{equation}
\tilde s_+ = v_{i} + \pi g \rho_{i} = v_{-} + \pi g \rho_{-}.
\label{e:vellead}
\end{equation}
Equating $s_+= \tilde s_+$, we obtain the velocity of the solitary wave edge of the original CS DSW, which, 
despite the presence of the expansion fan and the intermediate shelf, is equal to the characteristic velocity 
$V_+=r_+= v+ \pi g \rho$ (\ref{e:dless_r}) of the dispersionless CS equations evaluated at the left state $(\rho_-, u_-)$, 
as for the uni-directional BO equation (see (\ref{e:grouplinbobnew})). This velocity can be related to the amplitude of the 
soliton at the leading edge of the DSW through the soliton solution of the CS system (\ref{e:mass1}) and (\ref{e:mom1}), which 
is \cite{abanov2}
\begin{equation}
 \rho = \bar \rho + \frac{1}{\pi} \frac{a_{\rho}}{\theta^{2} + a_{\rho}^{2}}, \quad
 v = \bar v + g \frac{a_{v}}{\theta^{2} + a_{v}^{2}} , \quad  \theta = x - c_{s}t,
\label{e:sol}
\end{equation}
where $\bar \rho$ and $\bar v$ are the background state and the travelling phase, respectively.  The soliton amplitudes 
in $\rho$ and $v$ are, respectively,
\begin{equation}
 a_{\rho} = \frac{\pi g^{2} \bar \rho}{(c_{s} - \bar v)^{2} - \pi^{2}g^{2}\bar \rho^{2}}, \quad
a_{v} = \frac{g(c_{s} - v_{0})}{(c_{s} - \bar v)^{2} - \pi^{2}g^{2} \bar \rho^{2}} ,
\label{e:amprel}
\end{equation}
$c_s$ being the solitary wave velocity.  We denote the amplitude in $\rho$ of the lead solitary wave of the DSW as $A_{s}$. 
Then equating $c_s=s_+$ and $\bar \rho = \rho_+$, the lead solitary wave amplitude is
\begin{equation}
 A_{s} = \frac{(s_+ - v_{+})^{2}}{\pi^{2} g^{2}\rho_{+}} - \rho_{+}  = \frac{(v_- - v_{+}+\pi g \rho_-)^{2}}{\pi^{2} 
g^{2}\rho_{+}} - \rho_{+}.
\label{e:amplead}
\end{equation}
The DSW admissibility conditions \cite{el2,borereview,hoefer} require that $s_+>s_i$, which gives
\begin{equation}\label{adm1}
v_{-} + \pi g \rho_{-} > v_{+} + \pi g \rho_{+}.
\end{equation}
The other two admissibility conditions related to causality, $s_+>V_+(\rho_+, v_+)$ and $s_i<V_+(\rho_i, v_i)$, 
yielding the same inequality (\ref{adm1}).

This completes the modulation theory solution for Case (i).  We add that the expansion fan solution
(\ref{e:rhofan}) and (\ref{e:vfan}) is a weak solution of the dispersionless CS equations (\ref{e:massless}) and (\ref{e:momles}) 
exhibiting discontinuities in derivatives at the edges. These discontinuities are resolved in the full CS system by 
dispersive terms, either smoothly or by rapid linear oscillations, depending on the sign of the first derivative change 
across the weak discontinuity, see Figures \ref{f:casesrho}(a) and \ref{f:casesv}(a).  See also \cite{gur} for the detailed 
analysis of the similar configuration in the KdV case. 

A special case of this Case (i) solution is when there is a vacuum ahead of the DSW, so that $\rho_{+} = 0$. This case 
corresponds to the classical shallow water dam break problem whose solution in the dispersionless case does not contain 
shock waves \cite{whitham} and is described by a single expansion fan. Hence, for the dam break initial data we do not 
expect DSW formation in the full dispersive theory either.  Indeed, the dispersive hydrodynamic dam break problem was 
studied in \cite{elnls} in the framework of modulation theory for the defocusing NLS equation and was shown to produce 
the same single expansion wave as for classical shallow water theory.  

The absence of a DSW for $\rho_+=0$ may appear to contradict expression (\ref{e:amplead}) for the DSW amplitude at the leading 
edge, which apparently blows up in the limit $\rho_+ \to 0$.  This limit, however, is a singular one for the constructed 
modulation solution.
%  as  for $\rho_+ =0$ the solution does not contain intermediate shelf and DSW  and consists only of the expansion wave (\ref{e:rhofan}) and (\ref{e:vfan}). As a result $v_-$ and $v_+$ cease to be independent parameters as $\rho_+ \to 0$. 
Let $\rho_+ \ll 1$.  Since we expect that the intermediate shelf must disappear in the dam break limit, we assume that 
$\rho_i=\rho_+ + \delta$, where $\delta \sim \rho_+$.  Using (\ref{e:cp}) and (\ref{e:cm}) we readily obtain that 
$v_- - v_+ + \pi g \rho_-= 2 \pi g \delta$.  Substituting this into the lead wave amplitude (\ref{e:amplead}), 
we obtain $A_s \sim \rho_+$.  Hence, both the shelf and the DSW disappear in the limit $\rho_+ \to 0$, as 
expected.  This effect is obviously not captured by a unidirectional BO approximation of the CS Riemann problem 
\cite{abanov1}.

The solution for the CS hydrodynamics dam break problem then becomes a single expansion fan
\begin{equation}
 \rho = \left\{ \begin{array}{cc}
                \rho_{-}, & \frac{x}{t} < v_{-} - \pi g \rho_{-}, \\
                \frac{\pi g\rho_{-} + v_{-} - \frac{x}{t}}{2\pi g}, & v_{-} - \pi g \rho_{-} \le \frac{x}{t} \le
                v_{-} + \pi g \rho_{-}, \\
                0, &  v_{-} + \pi g \rho_{-} < \frac{x}{t}
               \end{array}
        \right. 
\label{e:rhodam}
\end{equation}
and
\begin{equation}
 v = \left\{ \begin{array}{cc}
                v_{-}, & \frac{x}{t} < v_{-} - \pi g \rho_{-}, \\
                \frac{x}{2t} + \frac{1}{2}\pi g \rho_{-} + \frac{1}{2}v_{-}, & v_{-} - \pi g \rho_{-} \le \frac{x}{t} \le
                v_{-} + \pi g \rho_{-}, \\
                v_{-} + \pi g \rho_{-}, &  v_{-} + \pi g \rho_{-} < \frac{x}{t}.
               \end{array}
        \right. 
\label{e:vdam}
\end{equation}
%We note the dispersive hydrodynamic dam break problem was studied in \cite{elnls} in the framework of the modulation theory for the defocusing NLS equation and was shown to produce the same classical single expansion wave.  

\subsection{Case (ii):  Two DSWs}
\label{s:case2}

%The expansion fan solution (\ref{e:rhofan}) and (\ref{e:vfan}) is valid if $v_{-} - \pi g\rho_{-} < v_{+} - \pi g\rho_{+}$
%and the DSW solution is valid if $v_{+} + \pi g \rho_{+} < v_{-} + \pi g \rho_{-}$.  
For  $v_{-} - \pi g\rho_{-} = v_{+} - \pi g\rho_{+}$ the expansion wave disappears and only the fast DSW is left.  Furthermore, 
if $v_{-} - \pi g\rho_{-} > v_{+} - \pi g\rho_{+}$, from the solution (\ref{e:inter}) for the intermediate level $\rho_{i}$ 
we have $\rho_{i} \ge \rho_{-}$ in this case.  The intermediate shelf then rises above the rear level $\rho_{-}$, so that a 
DSW must link $\rho_{i}$ to $\rho_{-}$.  Case (ii) then consists of two DSWs either side of the intermediate level 
$\rho_{i}, v_i$, propagating in opposite directions relative to the intermediate level $v_i$, as illustrated in 
Figures \ref{f:casesrho}(b) and \ref{f:casesv}(b).

The intermediate levels $\rho_{i}$ and $v_{i}$ remain the same, (\ref{e:inter}), as the Riemann invariant on $C_{-}$ is 
conserved through the fast DSW and the Riemann invariant on $C_{+}$ is conserved through the slow DSW.  
The fast DSW is then the same as that for case (i), with the leading, solitary wave edge given by (\ref{e:vellead}) 
and (\ref{e:amplead}) and the trailing, harmonic wave edge given by (\ref{e:grouplintrail}).  The slow DSW is given by 
the mirror image of the fast one.  In what follows we shall use the superscripts ``$s$''  and ``$f$'' for the quantities 
characterising slow and fast DSWs, respectively.  As already stated, the fast DSW is the same as above for case
(i).  Its trailing, linear edge is given by Eq.\ (\ref{e:grouplintrail}) and its leading, solitary wave edge by 
Eqs.\ (\ref{e:vellead}) and (\ref{e:amplead}).  Hence,
\begin{equation}
 k_i^{f} = 2\pi \left( \rho_{i} - \rho_{+} \right), \quad s_i^{f} = v_{+} + \pi g \rho_{+}
\label{e:linright2}
\end{equation}
and its leading, solitary wave edge has the velocity and amplitude
\begin{equation}
 s_+^{f} = v_{i} + \pi g \rho_{i} = v_{-} + \pi g\rho_{-}, \quad A_{s}^{f} = \frac{\left( s_+^{f} - v_{+}\right)^{2}}
{\pi^{2}g^{2}\rho_{+}} - \rho_{+} .
\label{e:solright2}
\end{equation}
The slow DSW is symmetric to the fast DSW.  The above solutions for the trailing, harmonic edge (\ref{e:grouplintrail})
and the leading, soliton edge (\ref{e:vellead}) and (\ref{e:amplead}) then give the trailing, linear and leading,
soliton edges of the slow DSW as
\begin{equation}
 k_i^{s}  = 2\pi \left( \rho_{i} - \rho_{-} \right), \quad s_i ^{s} = v_{-} - \pi g \rho_{-}
\label{e:linleft2}
\end{equation}
and
\begin{equation}
s_+^s  = v_{i} - \pi g \rho_{i} = v_{+} - \pi g \rho_{+}, \quad A_s^{s}  = \frac{\left( s_+^{s} - v_{-}\right)^{2}}
{\pi^{2}g^{2}\rho_{-}} - \rho_{-} ,
\label{e:solleft2}
\end{equation}
respectively.  It is noted that the edges of the slow DSW are in the correct order $s_+^s < s_i^s$ as 
$v_{-} - \pi g\rho_{-} > v_{+} - \pi g\rho_{+}$ and the fast DSW solution is valid if $v_{+} + \pi g\rho_{+}
< v_{-} + \pi g \rho_{-}$. Hence, the admissibility conditions for both DSWs are satisfied.  

There is another condition which ensures that the DSWs for case (ii) are fully realised and are not overlapping.
This condition is $s_i^{f}>s_i^{s}$, which translates to
\begin{equation}\label{overlap}
v_{+} + \pi g \rho_{+}  \ge v_{-} - \pi g \rho_{-}.
\end{equation}
Violation of condition (\ref{overlap}) leads to DSW interaction, which is a separate pattern described in case (vi) below.

\subsection{Case (iii):  Downstream propagating expansion wave and DSW}
\label{s:case3}

If $\rho_{-} < \rho_{+}$, then the DSW and expansion fan flip around, with the expansion fan to the right of
the (slow) DSW and the linear edge of the DSW to the right of the soliton edge, as illustrated in Figures \ref{f:casesrho}(c)
and \ref{f:casesv}(c).  In addition, the expansion fan is on the $C_{+}$ characteristic with the Riemann invariant on 
$C_{-}$ constant.  Finally, the Riemann invariant on the $C_{+}$ characteristic is constant through the slow DSW.
Repeating the derivation of the intermediate level as for case (i), we have that the intermediate level for this 
case (iii) is given by the same expressions (\ref{e:inter}).
%\begin{equation}
% \rho_{i} = \frac{1}{2} \left( \rho_{-} + \rho_{+} \right) + \frac{v_{-} - v_{+}}{2\pi g}, \quad
%v_{i} = \frac{\pi g}{2} \left( \rho_{-} - \rho_{+} \right) + \frac{1}{2} \left( v_{+} + v_{-} \right).
%\label{e:intercase3}
%\end{equation}
Using this intermediate level, the expansion fan on the characteristic $C_{+}$ (\ref{e:dless}) is
\begin{equation}
 \rho = \left\{ \begin{array}{cc}
                   \rho_{i}, & \frac{x}{t} < \pi g \rho_{-} + v_{-}, \\
                   \frac{1}{2 \pi g} \left[ \frac{x}{t} + \pi g \rho_{+} - v_{+} \right], & 
                   \pi g \rho_{-} + v_{-} \le \frac{x}{t} \le \pi g \rho_{+} + v_{+}, \\
                   \rho_{+}, & \frac{x}{t} > \pi g \rho_{+} + v_{+}
                \end{array}
        \right.
\label{e:fancase3}
\end{equation}
and
\begin{equation}
 v = \left\{ \begin{array}{cc}
                   v_{i}, & \frac{x}{t} < \pi g \rho_{-} + v_{-}, \\
                   \frac{x}{2t} - \frac{1}{2}\pi g \rho_{+} + \frac{1}{2} v_{+}, & 
                   \pi g \rho_{-} + v_{-} \le \frac{x}{t} \le \pi g \rho_{+} + v_{+}, \\
                   v_{+}, & \frac{x}{t} > \pi g \rho_{+} + v_{+}.
                \end{array}
        \right.
\label{e:fanvcase3}
\end{equation}
Since the intermediate level is the same as in case (ii), the trailing and leading edges of the slow DSW are given by the same 
Eqs.\ (\ref{e:linleft2}) and (\ref{e:solleft2}), respectively.

This solution for case (iii) holds if $v_{+} - \pi g \rho_{+} < v_{-} - \pi g \rho_{-}$ and $\pi g \rho_{-} + v_{-} 
< \pi g \rho_{+} + v_{+}$.  
%These inequalities certainly hold if $v_{+} = v_{-} = 0$.

\subsection{Case (iv):  Two expansion waves connected by a constant plateau}
\label{s:sect4}

The velocity of the leading, solitary wave edge of the DSW of case (i) is given by (\ref{e:vellead}) and that for the
trailing, linear wave edge is given by (\ref{e:grouplintrail}).  This DSW solution is then valid if 
$v_{-} + \pi g\rho_{-} > v_{+} + \pi g\rho_{+}$.  If 
\begin{equation}
 \pi g \rho_{-} - \pi g \rho_{+} \le v_{+} - v_{-},
\label{e:case4}
\end{equation}
the DSW disappears and only the expansion wave is left.  In fact, there are two expansion waves,
one linking $\rho_{-}$ to $\rho_{i}$ on the $C_{-}$ characteristic and the other linking $\rho_{i}$ to $\rho_{+}$
on the $C_{+}$ characteristic, as illustrated in Figures \ref{f:casesrho}(d) and \ref{f:casesv}(d).  The expansion wave 
linking $\rho_{-}$ to $\rho_{i}$ is given by (\ref{e:rhofan}), but with the intermediate shelf terminated at the right 
expansion fan.  Furthermore, $\rho_{i}$ and $v_{i}$ are still given by (\ref{e:inter}) due to this level linking the 
constant Riemann invariant on $C_{+}$ from the level behind $\rho_{-}$ and $v_{-}$ and the constant Riemann invariant 
on $C_{-}$ from the level ahead $\rho_{+}$ and $v_{+}$. 

The new expansion fan solution on $C_{+}$ linking the intermediate level $\rho_{i}$ and $v_{i}$ to the level ahead 
$\rho_{+}$ and $v_{+}$, using the Riemann invariant form of the dispersionless equations (\ref{e:massless}) and 
(\ref{e:momles}), is 
\begin{equation}
 \rho = \left\{ \begin{array}{cc}
                  \rho_{i}, & v_{+} - \pi g \rho_{+} < \frac{x}{t} < v_{-} + \pi g \rho_{-}, \\
                  \frac{\frac{x}{t} + \pi g\rho_{+} - v_{+}}{2\pi g}, & v_{-} + \pi g \rho_{-} \le \frac{x}{t} 
                  \le v_{+} + \pi g \rho_{+}, \\
                  \rho_{+}, & \frac{x}{t} > v_{+} + \pi g \rho_{+} ,
                \end{array}
        \right. 
\label{e:fancp}
\end{equation}
with the $v$ given by $v = v_{+} + \pi g (\rho - \rho_{+})$.  The expansion fan (\ref{e:rhofan}) and (\ref{e:vfan})
linking the intermediate level to the level behind $\rho_{-}$ can now be joined with this expansion fan to give the 
final expansion wave solution  
\begin{equation}
 \rho = \left\{ \begin{array}{cc}
                \rho_{-}, & \frac{x}{t} < v_{-} - \pi g \rho_{-}, \\
                \frac{\pi g\rho_{-} + v_{-} - \frac{x}{t}}{2\pi g}, & v_{-} - \pi g \rho_{-} \le \frac{x}{t} \le
                v_{+} - \pi g \rho_{+}, \\
                \rho_{i}, & v_{+} - \pi g \rho_{+} < \frac{x}{t} < v_{-} + \pi g \rho_{-}, \\
                \frac{\frac{x}{t} + \pi g\rho_{+} - v_{+}}{2\pi g}, & v_{-} + \pi g \rho_{-} \le \frac{x}{t} 
                  \le v_{+} + \pi g \rho_{+}, \\
                  \rho_{+}, & \frac{x}{t} > v_{+} + \pi g \rho_{+} 
               \end{array}
        \right. 
\label{e:rhocase4full}
\end{equation}
and
\begin{equation}
 v = \left\{ \begin{array}{cc}
                v_{-}, & \frac{x}{t} < v_{-} - \pi g \rho_{-}, \\
                \frac{x}{2t} + \frac{1}{2}v_{-} + \frac{1}{2}\pi g\rho_{-}, & v_{-} - \pi g \rho_{-} \le \frac{x}{t} \le
                v_{+} - \pi g \rho_{+}, \\
                v_{i}, & v_{+} - \pi g \rho_{+} < \frac{x}{t} < v_{-} + \pi g \rho_{-}, \\
                \frac{x}{2t} + \frac{1}{2}v_{+} - \frac{1}{2}\pi g \rho_{+}, & v_{-} + \pi g \rho_{-} \le \frac{x}{t} 
                  \le v_{+} + \pi g \rho_{+} \\
                  v_{+}, & \frac{x}{t} > v_{+} + \pi g \rho_{+} .
               \end{array}
        \right. 
\label{e:vcase4full}
\end{equation}

\subsection{Case (v):  Two expansion waves and a vacuum region}
\label{s:case5}

The intermediate shelf (\ref{e:inter}) can become negative if $\pi g(\rho_{+} + \rho_{-}) < v_{+} - v_{-}$.
As $\rho \ge 0$, this is not a valid solution.  Hence, a new solution form is needed for 
\begin{equation}
 \pi g\left( \rho_{+} + \rho_{-} \right) \le v_{+} - v_{-}.
\label{e:case5}
\end{equation}
In this regime the solution is obtained by cutting out the region in which $\rho < 0$ in the double expansion
wave solution of Section \ref{s:sect4}.  An example of this vacuum point solution is shown in Figures \ref{f:casesrho}(e)
and \ref{f:casesv}(e).  Hence, the solution in the region (\ref{e:case5}) is
\begin{equation}
 \rho = \left\{ \begin{array}{cc}
                \rho_{-}, & \frac{x}{t} < v_{-} - \pi g \rho_{-}, \\
                \frac{\pi g\rho_{-} + v_{-} - \frac{x}{t}}{2\pi g}, & v_{-} - \pi g \rho_{-} \le \frac{x}{t} \le
                v_{-} + \pi g \rho_{-}, \\
                0, & v_{-} + \pi g \rho_{-} < \frac{x}{t} < v_{+} - \pi g \rho_{+}, \\
                \frac{\frac{x}{t} + \pi g\rho_{+} - v_{+}}{2\pi g}, & v_{+} - \pi g \rho_{+} \le \frac{x}{t} 
                  \le v_{+} + \pi g \rho_{+}, \\
                  \rho_{+}, & \frac{x}{t} > v_{+} + \pi g \rho_{+} 
               \end{array}
        \right. 
\label{e:rhocase5full}
\end{equation}
and
\begin{equation}
 v = \left\{ \begin{array}{cc}
                v_{-}, & \frac{x}{t} < v_{-} - \pi g \rho_{-}, \\
                \frac{x}{2t} + \frac{1}{2}v_{-} + \frac{1}{2}\pi g\rho_{-}, & v_{-} - \pi g \rho_{-} \le \frac{x}{t} \le
                v_{-} + \pi g \rho_{-}, \\
                v_{i}, & v_{+} - \pi g \rho_{+} < \frac{x}{t} < v_{-} + \pi g \rho_{-}, \\
                \frac{x}{2t} + \frac{1}{2}v_{+} - \frac{1}{2}\pi g \rho_{+}, & v_{-} + \pi g \rho_{-} \le \frac{x}{t} 
                  \le v_{+} + \pi g \rho_{+}, \\
                  v_{+}, & \frac{x}{t} > v_{+} + \pi g \rho_{+} .
               \end{array}
        \right. 
\label{e:vcase5full}
\end{equation}

\subsection{Case (vi): Overlapping upstream and downstream DSWs}
\label{s:case6}

This case is similar to case (ii), exhibiting two counter-propagating DSWs, but there is the ``wrong'' ordering 
of the DSW trailing edge velocities, $s_{i}^{f} < s_i^{s}$, which violates the admissibility condition (\ref{overlap}) 
for case (ii), which implies that the two DSWs overlap and interact.  The solution of the Riemann problem for the 
CS equations then consists of two partially realised DSWs connected by an oscillating region due to their interaction.  Generally, 
this intermediate region must be described by a modulated two phase wave solution, but as was shown for the equivalent 
case for the defocusing NLS equation \cite{bik,elnls} the self-similarity requirement for the modulation solution 
of the Riemann problem leads to this region being described by a {\it non-modulated} single phase wave, an ``oscillating plateau'' 
with constant mean.  This solution for $\rho$  and $v$ is illustrated in Figures \ref{f:casesrho}(f) and 
\ref{f:casesv}(f) (note that the intermediate interaction region in the numerical solution exhibits visible modulations 
due to the deviation of the initial data from a sharp step in the numerical simulations).  As the DSWs are now only  
partial, full modulation theory is needed for their solution and the DSW fitting method does not apply as the trailing, 
harmonic edges of the DSWs are not realised.  This case will then not be considered further here.

%\section{Comparison with Numerical Solutions}
%\label{s:comp}
%
%\begin{figure}
%\centering
%\includegraphics[width=0.33\textwidth,angle=270]{case1ampcomp.eps}
%\includegraphics[width=0.33\textwidth,angle=270]{case1velcomp.eps}
%\caption{Comparisons between modulation theory and numerical solutions for Case 1.
%(a) Lead solitary wave amplitude $A_{s}$ of DSW.  Numerical solution: (blue) crosses; modulation theory:
%(blue) line, (b) Velocity of leading edge $s_{+}$ and trailing edge $c_{g}$ of DSW.  Numerical solution
%for $s_{+}$: upper (blue) crosses; modulation theory solution for $s_{+}$: upper (blue) line.  Numerical 
%solution for $c_{g}$: lower (green) stars; modulation theory solution for $c_{g}$: lower (green) dashed line. 
%The parameter values are $\rho_{-}=5$, $v_{+} = v_{-} = 0$ and $g=0.5$.}
%\label{f:case1comp}
%\end{figure}

\section{Comparison with Numerical Solutions}
\label{s:comp}

\begin{figure}
\centering
\includegraphics[width=0.33\textwidth,angle=270]{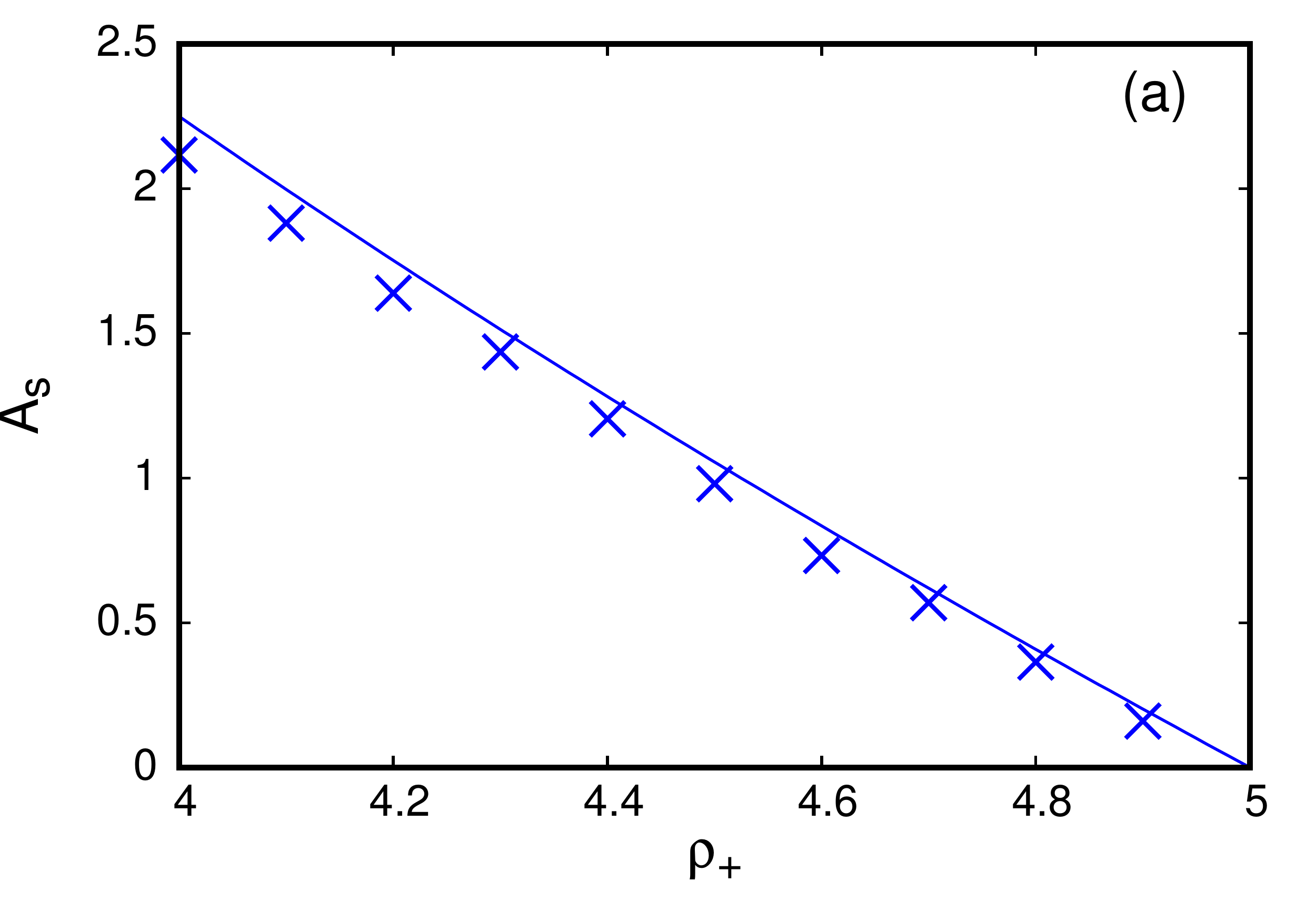}
\includegraphics[width=0.33\textwidth,angle=270]{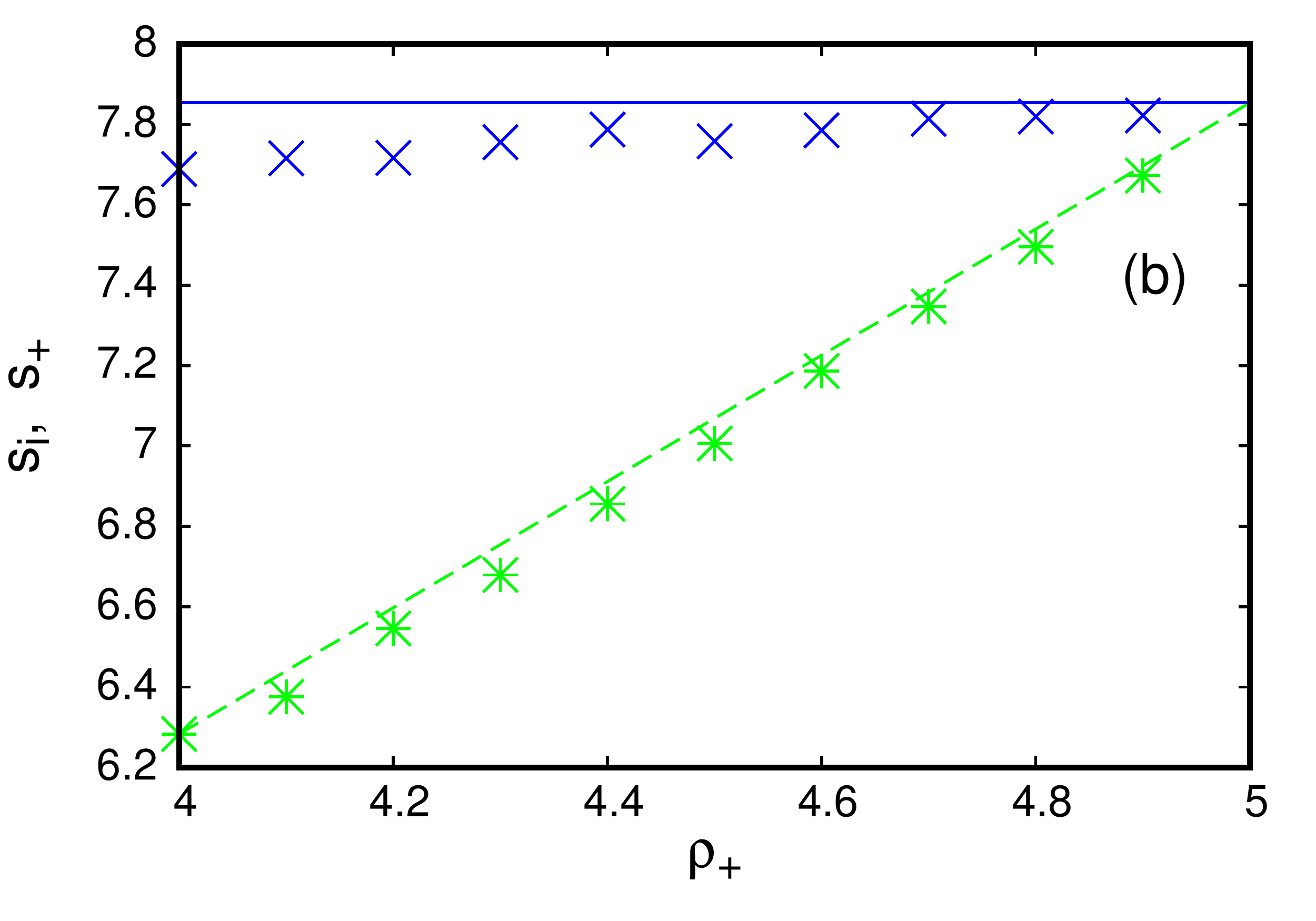}
\caption{Comparisons between modulation theory and numerical solutions for case (i).
(a) Lead solitary wave amplitude $A_{s}$ of DSW.  Numerical solution: (blue) crosses; modulation theory:
(blue) line, (b) Velocity of leading edge $s_{+}$ and trailing edge $s_i$ of DSW.  Numerical solution
for $s_{+}$: upper (blue) crosses; modulation theory solution for $s_{+}$: upper (blue) line.  Numerical 
solution for $s_{i}$: lower (green) stars; modulation theory solution for $s_{i}$: lower (green) dashed line. 
The parameter values are $\rho_{-}=5$, $v_{+} = v_{-} = 0$ and $g=0.5$.}
\label{f:case1comp}
\end{figure}

The modulation theory solutions of the previous section will now be compared with full numerical 
solutions of the CS system (\ref{e:mass1}) and (\ref{e:mom1}).  The Hilbert transform dispersive
term in these equations is most easily dealt with using a Fourier based method, so a pseudo-spectral
method based on that of Fornberg and Whitham \cite{bengt} will be used.  This classic pseudo-spectral 
method is then made more accurate by propagating forward in time $t$ using a fourth order Runge-Kutta
scheme and the stability is improved by propagating in $t$ in Fourier space rather than real space
\cite{chan,tref}.  However, even with these accuracy and stability improvements, it was found that
the time step $\Delta t$ needed to be $10^{-4}$ to $10^{-5}$ for a space step of order $\Delta x = 0.05$
for the scheme to be stable for jump heights $|\rho_{-} - \rho_{+}|$ near $1$.  Furthermore, the jump in the 
initial condition (\ref{e:rhoic}) was smoothed using a hyperbolic tangent to link the levels ahead and
behind in $\rho$ and $v$.  As a Fourier method was used to solve the CS system, in the numerical solutions
the step initial condition (\ref{e:rhoic}) was converted to a hat initial condition with $\rho = \rho_{+}$
and $v = v_{+}$ for $x < -L$ where $L$ is large.  In this manner, the numerical initial condition is
periodically continuous.  The parameter $L$ is chosen large enough so that the waves of interest generated
from $x=0$ do not interact with those generated from $x = -L$.

%\begin{figure}
% \centering
% \includegraphics[width=0.33\textwidth,angle=270]{compcase1mod.eps}
% \includegraphics[width=0.33\textwidth,angle=270]{compcase4mod.eps}
% \caption{Comparison of non-dispersive modulation solution and numerical solutions. (a) Case 1.  
% $\rho_{+}=4.5$, $\rho_{-}=5$, $v_{+}= 0$, $v_{-} = 0$, $t=100$, $g=0.5$, (b) Case 4.  $\rho_{+}=4.5$, 
% $\rho_{-}=5$, $v_{+}= 0$, $v_{-} = -1$, $t=20$, $g=0.5$, (c) Case 5.
% Numerical solution:  solid (red) line; modulation solution: dashed (green) line.}
%\end{figure}
\begin{figure}
 \centering
 \includegraphics[width=0.33\textwidth,angle=270]{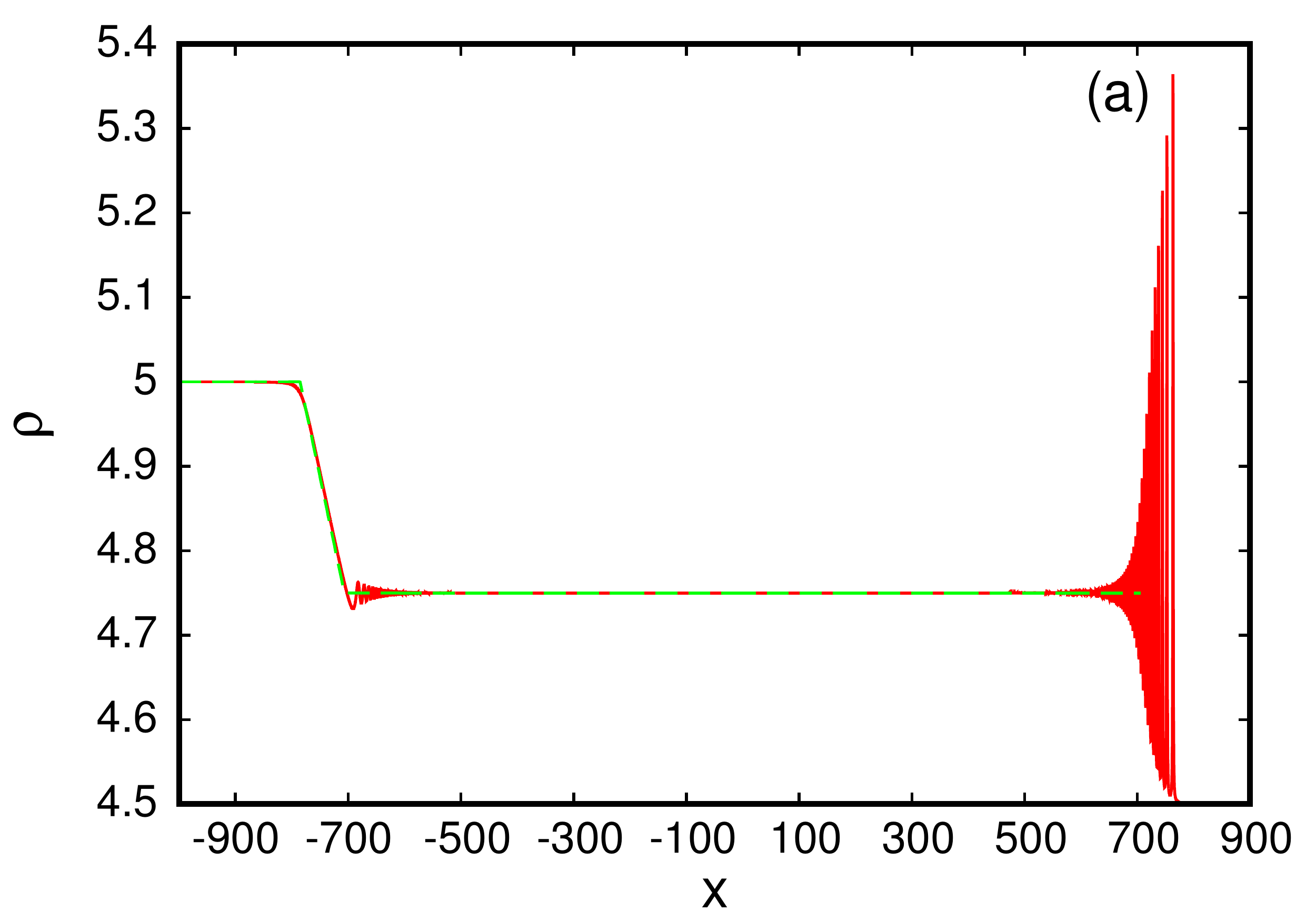}
 \includegraphics[width=0.33\textwidth,angle=270]{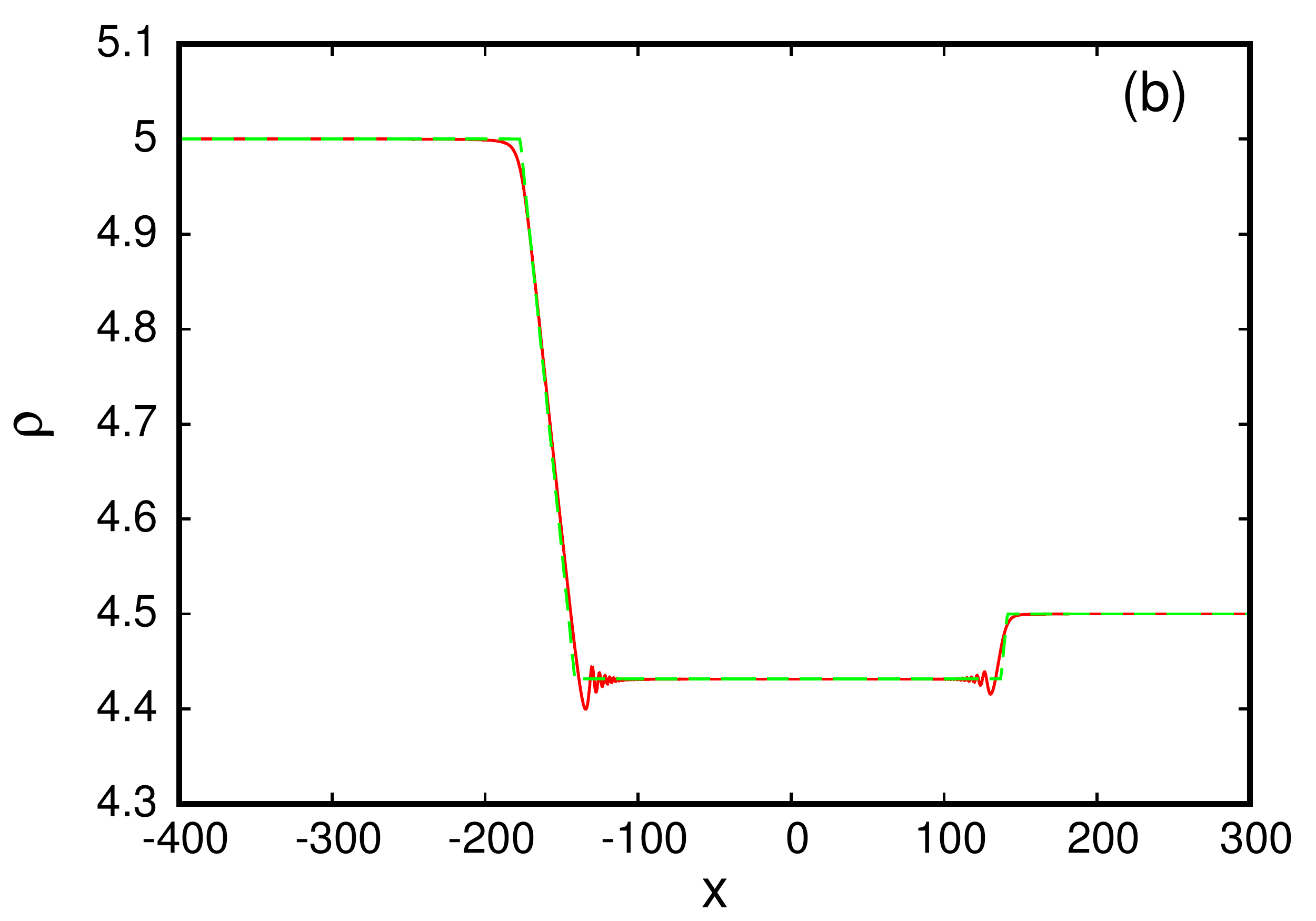}
 \includegraphics[width=0.33\textwidth,angle=270]{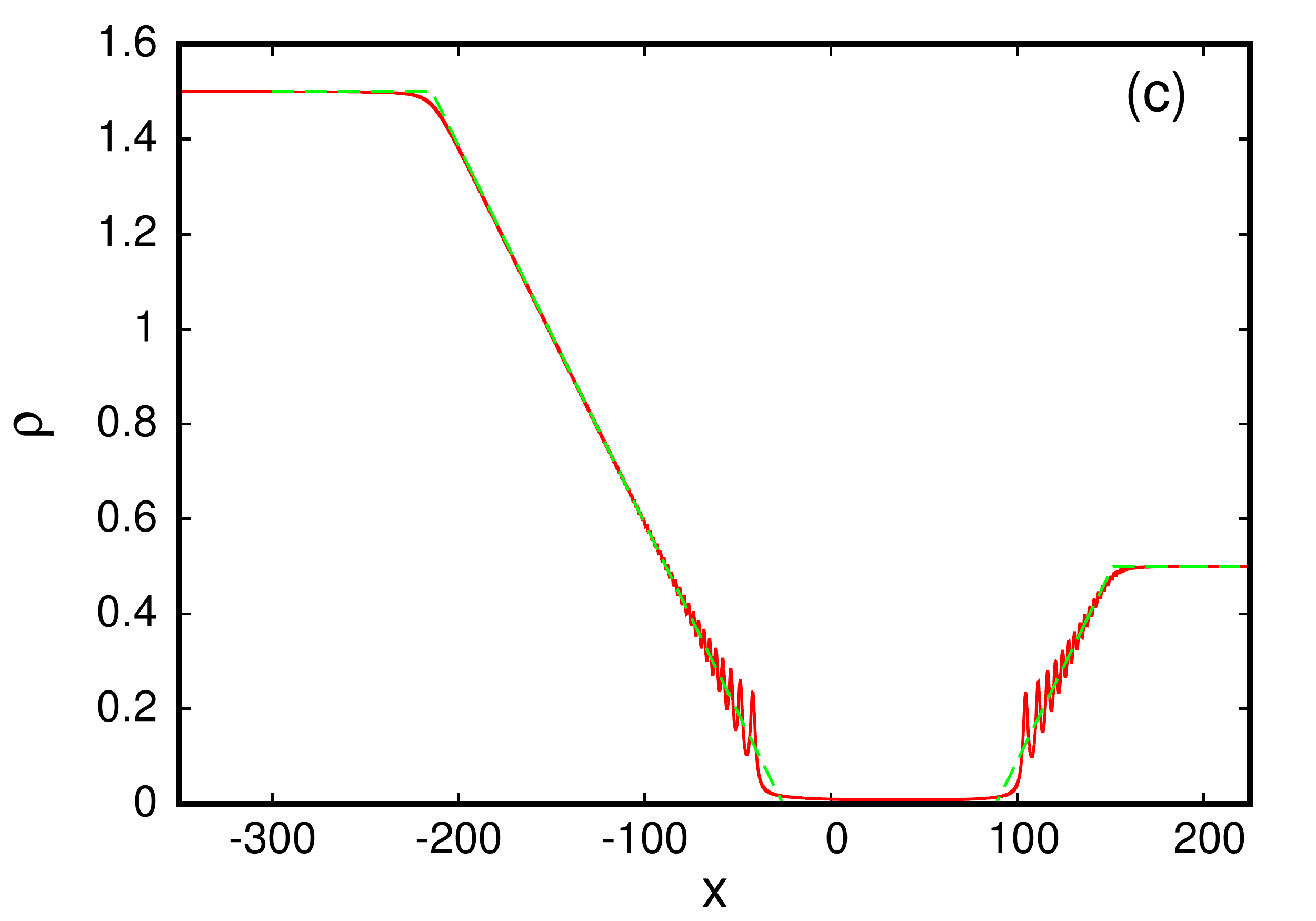}
 \caption{Comparison of non-dispersive modulation solution and numerical solutions. (a) case (i).  
 $\rho_{+}=4.5$, $\rho_{-}=5$, $v_{+}= 0$, $v_{-} = 0$, $t=100$, (b) case (iv).  $\rho_{+}=4.5$, 
 $\rho_{-}=5$, $v_{+}= 0$, $v_{-} = -1$, $t=20$, (c) case (v). $\rho_{+} = 0.5$, $\rho_{-}=1.5$, 
 $v_{+}=3$, $v_{-}=-3$, $t=40$.   Numerical solution:  solid (red) line; modulation solution: dashed 
 (green) line. $g = 0.5$.}
 \label{f:expandcomp}
\end{figure}

Figure \ref{f:case1comp} shows comparisons between the modulation theory solution for case (i) outlined in
Section \ref{s:case1} for a fast DSW and a slow expansion wave.  Shown in Figure \ref{f:case1comp}(a) is the
amplitude $A_{s}$ of the lead solitary wave of the DSW as a function of the level ahead $\rho_{+}$ for 
$v_{+}=0$ and fixed levels behind $\rho_{-}$ and $v_{-}$.  It can be seen that the modulation theory gives 
an excellent prediction for the lead wave amplitude.  The agreement is nearly perfect for low jump heights, 
with small deviations as the jump height is increased.  There is similar agreement for the velocity $s_{+}$ 
of the leading, solitary wave edge and $s_{i}$ of the trailing, linear wave edge of the DSW shown in Figure 
\ref{f:case1comp}(b).  The leading edge velocity $s_{+}$ shows the same variation in agreement with the level 
ahead $\rho_{+}$ as the lead amplitude $A_{s}$, with the agreement becoming slightly worse as the jump height 
increases, leading to higher lead wave amplitude.  The velocity $s_{i}$ of the trailing, linear edge shows uniform 
agreement with the jump height.  This contrast with the leading edge is because the trailing waves have 
a group velocity which is independent of amplitude, as for all linear waves.  

Figure \ref{f:expandcomp} shows comparisons for the non-dispersive portion of the modulation solution 
with numerical results for cases (i), (iv) and (v)  for some representative choices of the initial 
data \eqref{e:rhoic}.  These non-dispersive sections are given by the 
expansion wave solutions (\ref{e:rhofan}) and (\ref{e:vfan}) for case (i) consisting of a fast DSW
and a slow expansion fan, (\ref{e:rhocase4full}) and (\ref{e:vcase4full}) for case (iv) consisting of
two expansion fans and (\ref{e:rhocase5full}) and (\ref{e:vcase5full}) for case (v) which is the special 
case of case (iv) for which the intermediate shelf hits the vacuum.  The agreement for these 
non-dispersive portions is near perfect for all cases.  The numerical solutions show small dispersive
wavetrains at the corners of the expansion fans.  These act to smooth out the discontinuities in
derivatives \cite{bengt}.  These small dispersive wavetrains can be determined by using 
higher order matching asymptotic theories  \cite{leach}. 

  Similar comparisons were also done for cases (ii) and (iii), but we do not present the results 
here as these cases are qualitatively similar to case (i).  We note that case (iii) is just case (i) with the 
DSW and expansion wave reversed, as can be seen from Figures \ref{f:casesrho}(a) and (c) and \ref{f:casesv}(a) and (c).  
Case (ii) is two DSWs, both of the same form as for case (i).  We also do not 
present comparisons for case (vi) as its analytical description is beyond the scope of the present work, as 
explained in Sec.~\ref{s:case6}, due to the interaction of two DSWs.

\section{Conclusions}

In this work, the dispersive shock wave (DSW) fitting method \cite{el2,borereview,el1} developed previously 
for nonlinear dispersive wave equations of KdV type has been extended to equations with a nonlocal Benjamin-Ono 
(BO) type dispersion term involving the Hilbert transform. The method enables the determination of the key 
``observables'' of a DSW, namely, its locus, the velocities of the leading and trailing edges and the amplitude 
of the leading solitary wave, realised as an inherent part of the DSW structure. Importantly, the DSW fitting method  
is based on generic properties of periodic solutions of the governing nonlinear wave equation and its associated modulation 
(Whitham) system and, thus, does not have integrability of the wave equation and the related existence of a Riemann 
invariant form for the modulation system as a pre-requisite.  Thus, the method can be applied to many physically 
relevant equations which are typically not integrable via the inverse scattering transform.   The extension of the 
method to BO type equations was required as their periodic solutions differ substantially from those of
KdV-type equations, exhibiting algebraic, rather than exponential, decay in the solitary wave limit and resulting 
in the non-applicability of the original DSW fitting construction \cite{el2, el1} for the determination of the solitary 
wave edge.

The appropriate modification of the DSW fitting method was developed first for the generalised BO equation 
(\ref{e:gbo}) using the underlying symmetries of this equation.  In the particular case of the standard BO 
equation (\ref{e:bo}) the resulting solutions for the leading and trailing edges of the DSW, as well as the lead soliton 
amplitude of the DSW, were found to agree exactly with the previously known full modulation DSW solution of 
this equation \cite{tim,matsuno1} derived from its Whitham modulation equations \cite{dob}.  The same extension 
of the DSW fitting method utilising the equation's underlying symmetries was then successfully applied to the bi-directional 
Calogero-Sutherland (CS) system (\ref{e:mass1}) and (\ref{e:mom1}), which has the form of Eulerian dispersive 
hydrodynamics with the equation of state $P(\rho) \sim \rho^3$ and the dispersion operator combining NLS and 
BO dispersion.  The developed extension enabled the construction of the full classification of solutions of
the basic Riemann step problem for the CS system,  which contains six possible solution forms, each representing 
a certain combination of two waves, DSWs and/or rarefaction waves, separated by a constant or oscillatory shelf. 

Although the CS system (\ref{e:mass1}) and (\ref{e:mom1}) can be viewed as a bi-directional BO equation \cite{abanov2},  
the constructed Riemann problem classification shows that some of its solutions cannot be interpreted (even qualitatively) 
in terms of the solutions of the BO equation (\ref{e:bo}).  Additionally, although the CS system is integrable, the 
developed construction does not make use of this integrability and, hence, admits a straightforward generalisation to 
non-integrable Eulerian dispersive hydrodynamics with an arbitrary convex pressure law $P(\rho)$ and a dispersion 
operator of CS type.

Thus, the class of nonlinear dispersive wave equations whose DSW closure can be constructed has been extended. This 
extension is important as equations with Benjamin-Ono dispersion govern wave phenomena in nature \cite{clarke,anne}.

\section*{Acknowledgments}

The work of L.T.K.N was supported by the Ruhr University Research School PLUS, funded by Germany's Excellence 
Initiative [DFG GSC 98/3].


\section*{References}

\begin{thebibliography}{99}

\bibitem{abanov2}  A.G. Abanov, E. Bettelheim and P. Wiegmann, ``Integrable hydrodynamics of Calogero-Sutherland
model: bidirectional Benjamin-Ono equation,'' {\em J. Phys.\ A: Math.\ Theor.,} {\bf 42}, 135201 (2009).

\bibitem{ablowitz}   M.J. Ablowitz and  H. Segur, {\it Solitons and inverse scattering transform},
SIAM Philadelphia (1981).

\bibitem{ablowitz_2017}  M. Ablowiz, G. Biondini and Q. Wang, Whitham modulation theory for the two-dimensional Benjamin-Ono equation, {\em Phys. Rev. E} {\bf 96}, 032225 (2017).

\bibitem{colloid}  X. An, T.R. Marchant and N.F. Smyth, ``Optical dispersive shock waves in defocusing colloidal 
media,'' {\em Physica D,} {\bf 342}, 45--56 (2017).

\bibitem{fleischer2}  C. Barsi, W. Wan, C. Sun and J.W. Fleischer, ``Dispersive shock waves with nonlocal nonlinearity,''  
{\em Opt.\ Lett.,} {\bf 32}, 2930--2932 (2007).

\bibitem{benjamin}  T.B. Benjamin, ``Internal waves of permanent form in fluids of great depth,'' 
{\em J. Fluid Mech.,} {\bf 29}, 559--562 (1967).

\bibitem{abanov1}  E. Bettelheim, A. G. Abanov and P. Wiegmann, ``Nonlinear quantum shock waves in fractional 
quantum Hall edge states,'' {\em Phys.\ Rev.\ Lett.,} {\bf 97}, 246401 (2006).

\bibitem{bik} R.E. Bikbaev, ``Finite-gap attractors and transition processes of the shock wave type in integrable systems,''
{\em Zap.\ Nauch.\ Semin.\ POMI,} {\bf 199}, 25--36 (1992).

\bibitem{bona} J.L. Bona and H. Kalisch, ``Singularity formation in the generalized Benjamin-Ono equation,'' 
{\em Discr.\ Cont.\ Dyn.\ Syst.,} {\bf 11}, 27--45 (2004). 

\bibitem{cal2}  F. Calogero, ``Solution of the one-dimensional $N$-body problems with quadratic and/or 
inversely quadratic pair potentials,'' {\em J. Math.\ Phys.,} {\bf 12}, 419--436 (1971).

\bibitem{clarke}  R.H. Clarke, R.K. Smith and D.G. Reid, ``The morning glory of
the Gulf of Carpentaria: an atmospheric undular bore,'' {\em Monthly
Weather Rev.,} {\bf 109}, 1726--1750 (1981).

\bibitem{chan}  T.F. Chan and T. Kerkhoven, ``Fourier methods with extended stability intervals for KdV,''
{\em SIAM J. Numer. Anal.,} {\bf 22}, 441--454 (1985).

\bibitem{trilloresfour}  M. Conforti and S. Trillo, ``Radiative effects driven by shock waves in cavity-less
four-wave mixing combs,'' {\em Opt.\ Lett.,} {\bf 39}, 5760--5763 (2014).

\bibitem{trillores}  M. Conforti, F. Baronio and S. Trillo, ``Resonant radiation shed by dispersive
shock waves,'' {\em Phys.\ Rev.\ A,} {\bf 89}, 013807 (2014).

\bibitem{trilloresnature}  M. Conforti, S. Trillo, A. Mussot and A. Kudlinski, ``Parametric excitation of 
multiple resonant radiations from localized wavepackets,'' {\em Sci.\ Rep.,} {\bf 5}, 1--5 (2015).

\bibitem{hnls}  M. Crosta, S. Trillo and A. Fratalocchi, ``The Whitham approach to dispersive shocks in systems with 
cubic-quintic nonlinearities,'' {\em New J. Phys.,} {\bf 14}, 093019 (2012).

\bibitem{dob} S. Yu. Dobrokhotov and I.M. Krichever, ``Multiphase solutions of the Benjamin-Ono Equation and  
their averaging,'' translated from {\em Matematicheskie Zametki,} {\bf 49}, 42--50 (1991).

\bibitem{el2}  G.A. El, ``Resolution of a shock in hyperbolic systems modified by 
weak dispersion,'' {\em Chaos,} {\bf 15}, 037103 (2005).

\bibitem{borereview}  G.A. El and M.A. Hoefer, ``Dispersive shock waves and modulation theory,'' {\em Physica D,}
{\bf 333}, 11--65 (2016).

\bibitem{siam_review} G.A. El, M.A. Hoefer and M. Shearer, ``Dispersive and diffusive-dispersive shock waves for 
nonconvex conservation laws,'' {\em SIAM Review,} {\bf 59}, 3--61 (2017).

\bibitem{el_krylov95}  G.A. El and A.V. Krylov, ``General solution of the Cauchy problem for the defocusing 
NLS equation in the Whitham limit,'' {\em Phys.\ Lett.\ A}, {\bf 203}, 77--82 (1995). 

\bibitem{elphoto}  G.A. El, A. Gammal, E.G. Khamis, R.A. Kraenkel and A.M. Kamchatnov, ``Theory of optical dispersive 
shock waves in photorefractive media,'' {\em Phys.\ Rev.\ A,} {\bf 76}, 0523813 (2007).

\bibitem{elnls}  G.A. El, V.V. Geogjaev, A.V. Gurevich and A.L. Krylov, ``Decay of an initial discontinuity in the 
defocusing NLS hydrodynamics,'' {\em Physica D,} {\bf 87}, 186--192 (1995).

\bibitem{kb_sapm2} G.A. El, R.H.J. Grimshaw and A.M. Kamchatnov, ``Wave breaking and the generation of undular bores 
in an integrable shallow-water system,'' {\em Stud.\ Appl.\ Math.,} {\bf 114}, 395--411 (2005).
 
\bibitem{kb_sapm1} G.A. El, R.H.J. Grimshaw and M.V. Pavlov, ``Integrable shallow-water equations and undular bores,'' 
{\em Stud.\ Appl.\ Math.,} {\bf 106}, 157--186 (2001).

\bibitem{el1} G.A. El, V.V. Khodorovskii and A.V. Tyurina, ``Determination of boundaries of unsteady oscillatory zone 
in asymptotic solutions of non-integrable dispersive wave equations,'' {\em Phys.\ Lett.\ A,} {\bf 318}, 526--536 (2003).

\bibitem{nemboreel}  G.A. El and N.F. Smyth, ``Radiating dispersive shock waves in non-local optical media,'' 
{\em Proc.\ Roy.\ Soc.\ London A,} {\bf 472}, 20150633 (2016).

\bibitem{trillo7}  J. Fatome, C. Finot, G. Millot, A. Armaroli and S. Trillo, 
``Observation of optical undular bores in multiple four-wave mixing,'' {\em Phys.\ Rev.\ X,} {\bf 4}, 021022 (2014).

\bibitem{flash}  H. Flaschka, M.G. Forest and D.W. McLaughlin, 1980, ``Multiphase averaging and the inverse 
spectral solution of the Korteweg-de Vries equation,'' {\em Comm.\ Pure Appl.\ Math.,} {\bf 33}, 739--784.

\bibitem{bengt}  B. Fornberg and G.B. Whitham, ``Numerical and theoretical study of certain non-linear wave 
phenomena,'' {\em Phil.\ Trans.\ Roy.\ Soc.\ Lond.\ Ser.\ A--- Math.\ and Phys.\ Sci.,} {\bf 289}, 373--404 (1978).

\bibitem{BO_trillo} J. Garnier, G. Xu, S. Trillo and A. Picozzi, ``Incoherent dispersive shocks in the spectral 
evolution of random waves,'' {\em Phys.\ Rev.\ Lett.,} {\bf 111}, 113902 (2013).

\bibitem{trillo6}  N. Ghofraniha, C. Conti, G. Ruocco and S. Trillo, ``Shocks in nonlocal media,'' {\em Phys.\
Rev.\ Lett.,} {\bf 99}, 043903 (2017).

\bibitem{hoefer} M. A. Hoefer, ``Shock waves in dispersive Eulerian fluids,'' {\em J. Nonlin.\ Sci.,} {\bf 24}, 
525--577 (2014).

\bibitem{gershenzon} A.V. Gurevich, N.I. Gershenzon, A.L. Krylov and N.G. Mazur, ``Solutions of the sine-Gordon 
equation by the modulated-wave method and application to a two-state medium,'' {\em Sov.\ Phys.\ Doklady,} {\bf 34}, 
246--248 (1989).

\bibitem{gur_meshch} A.V. Gurevich and A.P. Meshcherkin, ``Expanding self-similar discontinuities and shock waves 
in dispersive hydrodynamics,'' {\em Sov.\ Phys.\ JETP,} {\bf 60}, 732--740 (1984).

\bibitem{gur_kryl} A.V. Gurevich and A.L. Krylov, ``Dissipationless shock waves in media with positive dispersion,'' 
{\em Sov.\ Phys.\ JETP,} {\bf 65}, 944--953 (1987).

\bibitem{gur}  A.V. Gurevich and L.P. Pitaevskii,  ``Nonstationary structure of a collisionless shock wave,''
{\em Sov.\ Phys.\ JETP,} {\bf 33}, 291--297 (1974).
 
\bibitem{hoefer_euler} M.A. Hoefer, ``Shock waves in dispersive Eulerian fluids,'' {\em J. Nonlin.\ Sci.,} 
{\bf 24}, 525--577 (2014).
  
\bibitem{tim}  M.C. Jorge, A.A. Minzoni and N.F. Smyth, ``Modulation solutions for the Benjamin-Ono 
equation,'' {\em Physica D,} {\bf 132}, 1--18 (1999).

\bibitem{kamch_book}  A.M. Kamchatnov, {\em Nonlinear Periodic Waves and
their Modulations--- an Introductory Course}, World Scientific, Singapore (2000).

\bibitem{gardner}  A.M. Kamchatnov, Y.-H. Kuo, T.-C. Lin, T.-L. Horng, S.-C. Gou, R. Clift, G.A. El and 
R.H.J. Grimshaw, ``Undular bore theory for the Gardner equation,'' {\em Phys.\ Rev.\ E,} {\bf 86}, 036605 (2012).

%\bibitem{conduit1} N.K. Lowman and M.A. Hoefer, Dispersive shock waves in viscously
%deformable media, Journ. Fluid. Mech. {\bf 718},  524-557 (2013).

\bibitem{lax_lev} P.D. Lax, C.D. Levermore, ``The small dispersion limit of the Korteweg-de
Vries equation'' 1--3, {\em Comm.\ Pure Appl.\ Math.,} {\bf 36}, 253--290, 571--593, 809--830 (1983).

\bibitem{leach} J. A. Leach, D. J. Needham, ``The large-time development of the solution to an initial-value 
problem for the Korteweg-de Vries equation: I. Initial data has a discontinuous expansive
step,'' {\em Nonlinearity,} {\bf 21 (10)}, 2391--2408 (2008).

\bibitem{hoefer_conduit}  N.K. Lowman and M.A. Hoefer, ``Dispersive shock waves in viscously deformable media,'' 
{\em J. Fluid Mech.,} {\bf 718}, 524--557 (2013).

\bibitem{marchant_mkdv}  T.R. Marchant, ``Undular bores and the initial-boundary value problem for the modified 
Korteweg-de Vries equation,'' {\em Wave Motion,} {\bf 45,} 540--555 (2008).

\bibitem{matsuno1}  Y. Matsuno, ``The small dispersion limit of the Benjamin-Ono equation and the evolution of a 
step initial condition,'' {\em J. Phys.\ Soc.\ Japan,} {\bf 67}, 1814--1817 (1998).
 
\bibitem{matsuno2} Y. Matsuno, ``Nonlinear modulation of periodic waves in the small dispersion limit of the 
Benjamin-Ono equation,'' {\em Phys.\ Rev.\ E,} {\bf 58}, 7934--7940 (1998).

\bibitem{matsuno3} Y. Matsuno, V.S. Shchesnovich, A.M. Kamchatnov and R.A. Kraenkel,
``Whitham method for the Benjamin-Ono-Burgers equation and dispersive shocks'' {\em Phys.\ Rev.\ E,} {\bf 75}, 
016307 (2007).

\bibitem{miller_review} P.D. Miller, ``On the generation of dispersive shock waves,'' {\em Physica D,}
{\bf 333}, 66--83 (2016).

\bibitem{miller1}  P.D. Miller and A.N. Wetzel, ``The scattering transform for the Benjamin-Ono equation in the 
small-dispersion limit,'' {\em Physica D,} {\bf 333}, 185--199 (2016).

\bibitem{miller2}  P.D. Miller and Z. Xu, ``On the zero dispersion limit of the Benjamin-Ono Cauchy problem for 
positive initial data,'' {\em Comm.\ Pure Appl.\ Math.,} {\bf 64}, 205--270 (2011).

%\bibitem{sineg}  A.A. Minzoni and N.F. Smyth,  A modulation solution of the
%     signalling problem for the equation of self-induced transparency in the
%     Sine-Gordon limit,  Meth.\ Appl.\ Anal., {\bf 4}, 1--10 (1997).

\bibitem{sgbore}  A.A. Minzoni and N.F. Smyth, ``A modulation solution of the
     signalling problem for the equation of self-induced transparency in the
     Sine-Gordon limit,'' {\em Meth.\ Applic.\ Anal.,} {\bf 4}, 1--10 (1997).
     
\bibitem{novikov}  S.P. Novikov, S.V. Manakov, L.P. Pitaevskii and V.E. Zakharov, {\it Theory of Solitons: 
The Inverse Scattering Method}, Springer (1984).

% \bibitem{newell}  A.C. Newell, {\em Solitons in Mathematics and Physics,} SIAM, Philadelphia (1985).

%\bibitem{conduit}  N.K. Lowman and M.A. Hoefer, ``Dispersive hydrodynamics in viscous fluid conduits,''
%{\em Phys.\ Rev.\ E,} {\bf 88}, 023016 (2013).

\bibitem{ono}  H. Ono, ``Algebraic solitary waves in stratified fluids,'' {\em J. Phys.\ Soc.\ Jap.,}
{\bf 39}, 1082--1091 (1975).

\bibitem{pelin} D.E. Pelinovsky, ``Intermediate nonlinear Schr\"odinger equation for internal waves in 
a fluid of finite depth,'' {\em Phys.\ Lett.\ A,} {\bf 197,} 401--406 (1995).

\bibitem{polychronakos} A.P. Polychronakos, ``Waves and solitons in the continuum limit of the Calogero-Sutherland 
model,'' {\em Phys.\ Rev.\ Lett.,} {\bf 74,} 5153--5157 (1995).

\bibitem{anne}  V.A. Porter and N.F. Smyth, ``Modelling the Morning Glory of
     the Gulf of Carpentaria,'' {\em J. Fluid Mech.,} {\bf 454}, 1--20 (2002).

%\bibitem{cal1}  F. Calogero, ``Solution of a three-body problem in one dimension,'' {\em J. Math.\ Phys.,} 
%{\bf 10}, 2191--2197 (1969).

\bibitem{scott1}  D.R. Scott and D.J. Stevenson, ``Magma solitons,'' {\em Geophys.\ Res.\ Lett.,} {\bf 11}, 1161--1164 (1984).

\bibitem{nembore}  N.F. Smyth, ``Dispersive shock waves in nematic liquid crystals,'' {\em Physica D,} {\bf 333}, 
301--309 (2016).

\bibitem{nwshelf}  N.F. Smyth and P.E. Holloway, ``Hydraulic jump and undular bore
formation on a shelf break,'' {\em J. Phys.\ Ocean.,} {\bf 18}, 947--962 (1988).

\bibitem{hoefer_kawahara}  P. Sprenger and M.A. Hoefer, ``Shock waves in dispersive hydrodynamics with 
nonconvex dispersion,'' {\em SIAM J. Appl.\ Math.,} {\bf 77}, 26--50 (2017).

\bibitem{suth1}  B. Sutherland, ``Exact results for a quantum many-body problem in one dimension,''  
{\em Phys.\ Rev.\ A,} {\bf 4}, 2019--2021 (1971).

\bibitem{suth2}  B. Sutherland, ``Exact results for a quantum many-body problem in one dimension II,'' 
{\em Phys.\ Rev.\ A,} {\bf 5}, 1372--1376 (1972).

%\bibitem{suth3}  B. Sutherland, ``Exact ground-state wave function for a one-dimensional plasma,''
%{\em Phys.\ Rev.\ Lett.,} {\bf 34}, 1083--1084 (1975).

\bibitem{tref}  L.N. Trefethen, {\em Spectral Methods in MATLAB}, SIAM, Philadephia (2000).

\bibitem{fleischer}  W. Wan, S. Jia and J.W. Fleischer, ``Dispersive superfluid-like shock waves in nonlinear
optics,'' {\em Nature Phys.,} {\bf 3}, 46--51 (2007). 

\bibitem{boreexp}  J. Wang, J. Li, D. Lu, Q. Guo and W. Hu, ``Observation of surface
dispersive shock waves in a self-defocusing medium,'' {\em Phys.\ Rev.\ A,} {\bf 91}, 063819 (2015).

\bibitem{jfmmod}  G.B. Whitham, ``A general approach to linear and non-linear dispersive waves using a Lagrangian,''
{\em J. Fluid Mech.,} {\bf 22}, 273--283 (1965).

\bibitem{modproc}  G.B. Whitham, ``Non-linear dispersive waves,'' {\em Proc.\ Roy.\ Soc.\ London Ser.\ A,}
{\bf 283}, 238--261 (1965).

\bibitem{whitham}  G.B. Whitham, {\em Linear and Nonlinear Waves,} J. Wiley and Sons, New York (1974).

\bibitem{trillo2017} G. Xu, M. Conforti, A. Kudlinski, A. Mussot and S. Trillo, ``Dispersive dam-break flow of a 
photon fluid'', arXiv:1703.09019 [physics.optics] (2017)

\end{thebibliography}
\end{document}